\documentclass[conference]{IEEEtran}

\usepackage{graphicx}
\usepackage{booktabs}
\usepackage{amsmath}
\usepackage{multirow}
\usepackage{url}
\usepackage{hyperref}
\usepackage[caption=false,font=footnotesize,captionskip=1pt,farskip=3pt]{subfig}
\usepackage{makecell}
\usepackage{xcolor} 
\usepackage{float} 
\usepackage{pdflscape} 
\usepackage{tabularx}

\usepackage{cite}

\begin{document}

\title{Better Call Graphs: A New Dataset of Function Call Graphs for Malware Classification}

\author{
    \IEEEauthorblockN{Jakir Hossain, Jue Guo, Gurvinder Singh, Lukasz Ziarek, Ahmet Erdem Sar{\i}y\"{u}ce}
    \IEEEauthorblockA{University at Buffalo, Buffalo, NY, USA\\
        \{mh267, jueguo, gurvind2, lziarek, erdem\}@buffalo.edu}
}


\maketitle

\begin{abstract}
    Function call graphs (FCGs) have emerged as a powerful abstraction for malware detection, capturing the behavioral structure of applications beyond surface-level signatures. Their utility in traditional program analysis has been well established, enabling effective classification and analysis of malicious software. In the mobile domain, especially in the Android ecosystem, FCG-based malware classification is particularly critical due to the platform’s widespread adoption and the complex, component-based structure of Android apps.
    However, progress in this direction is hindered by the lack of large-scale, high-quality Android-specific FCG datasets. Existing datasets are often outdated, dominated by small or redundant graphs resulting from app repackaging, and fail to reflect the diversity of real-world malware. These limitations lead to overfitting and unreliable evaluation of graph-based classification methods.
    To address this gap, we introduce Better Call Graphs (BCG), a comprehensive dataset of large and unique FCGs extracted from recent Android application packages (APKs). BCG includes both benign and malicious samples spanning various families and types, along with graph-level features for each APK. Through extensive experiments using baseline classifiers, we demonstrate the necessity and value of BCG compared to existing datasets.
BCG is publicly available at {\bf\url{https://erdemub.github.io/BCG-dataset}}.

\end{abstract}


\section{Introduction}
Malware detection is a key task in the field of cybersecurity, especially within the Android ecosystem.
In most malware samples, minor changes in the source code of the original malware can lead to substantially different compiled code (e.g., through instruction reordering, branch inversion, and register allocation)~\cite{bayer06}. This is often exploited to bypass signature-based detection, a common method of malware detection~\cite{scott2017signature}. However, these minor source code modifications have little impact on the executable's control flow, which can be depicted using a function call graph (FCG) where functions are the nodes and the call relations are the edges. While FCGs have long been used in general program analysis, Android applications require specialized extraction and analysis methods due to their unique characteristics, such as lifecycle components, inter-component communication, and platform-specific APIs \cite{arzt14,li15,octeau16}.
Hence, FCG-based malware detection has been an important field of study, particularly within the realm of Android~\cite{ye17, freitas20}.

One challenge in FCG-based Android malware detection is that achieving accurate and robust classification models has been hampered by the limited availability of modern and representative large-scale Android-specific datasets.
Existing datasets typically contain old Android application packages (APKs) along with their corresponding FCGs.
Considering the dynamic landscape of Android ecosystem, obsolete APKs offer no benefit at all as they are developed on earlier version of Android.
In addition, the complexity of both benign and malicious Android applications has drastically changed in recent years.
Furthermore, most existing datasets contain many duplicate APKs, packaged with trivial differences, hence have a different name but the same FCG structure.
These limitations result in overfitting and unreliable evaluation of graph-based classification methods, as many approaches perform well on existing datasets but fail to generalize to real-world scenarios.
This performance degradation over time is related to \textit{concept drift}, where the characteristics of malware evolve, causing models trained on older data to become less effective~\cite{tripathi25}.

In this paper, we introduce a new and comprehensive dataset named Better Call Graphs (BCG), comprising extensive and distinct Android-specific function call graphs extracted from recent APKs.
The dataset includes benign samples as well as malware samples spanning various types and families.
We establish the necessity of BCG through the evaluation of several baseline approaches on existing datasets.
We show that existing datasets often yield misleading scores when state-of-the-art classifiers are applied.
Our dataset also contains graph-level APK features, capturing both the structural and behavioral characteristics of Android applications.
BCG is publicly available at {\bf{\url{https://erdemub.github.io/BCG-dataset}}}. 
The dataset is released under a CC-BY license, enabling free sharing and adaptation for research or development purposes.

In the rest of the paper, we first give a background and summarize related works on FCG-based Android malware detection and Android-based FCG datasets in Section~\ref{sec:relwork}.
Then we provide a detailed description of how the BCG is collected and filtered in Section~\ref{sec:Dataset_Collection}.
Next, we explain the properties of BCG in detail (Section \ref{sec:BCG_properties}) and perform graph classification experiments on our dataset, as well as other established datasets, using state-of-the-art graph classification methods (Section \ref{sec:experimental_evaluation}) --- we perform malware type as well as family classification.
Finally, we summarize our work and discuss the limitations in Section~\ref{sec:conc}.

\section{Related Work}
\label{sec:relwork}
The literature is rich with studies on Android malware detection by using various types of graphs such as Function Call Graphs (FCGs), Control Flow Graphs, and Network Flow Graphs.
These studies rely on existing malware datasets to evaluate the effectiveness of their methods.
Here, we review prior research on FCG-based detection for Android apps and discuss commonly used malware datasets.

\subsection{FCG-based Android Malware Detection}
Android apps, packaged in APK files, bundle all their components --- code, resources, and a manifest file.
Extracting various features from the code allows researchers to analyze how the app works and identify potential security risks.
FCGs, in particular, have proven valuable for malware classification by revealing how the app's functions interact, potentially exposing malicious behavior.
Researchers have actively explored diverse methods that leverage FCGs to analyze Android apps for security purposes {\cite{zhu18,feng20,cai21,xu21,vinayaka21}}.
These methods typically involve constructing FCGs and enriching them with node features, which can be either basic properties or more complex embeddings learned through graph neural networks.
The enriched FCGs are then used for identifying and classifying malware.
Classifications are typically performed to identify both the malware type and the malware family.
Malware type refers to the broad category of malicious behavior exhibited by a malware program. Common types of malware include Virus, Riskware, Trojan Horse, and Adware.
Malware family refers to a group of malware programs that share similar characteristics, codebase, or functionality, and are often named by antivirus companies based on their unique features.
E.g., within the "Trojan Horse" type, there might be the "Emotet" family, known for email spam and credential theft.

    {{MAMADroid and APIGraph utilized API semantics features to capture the semantic similarities between malware variants and analyze information flow for malware detection~\cite{onwuzurike19,zhang20}.
                In contrast,  Yuan et al. \cite{yuan20} focused on byte-level classification by converting malware binaries into Markov images and applying deep learning for detection. Meanwhile,  Fan et al. \cite{fan18} proposed a family-level classification approach by leveraging frequent subgraphs within FCGs.}}
 Zhu et al. \cite{zhu18} constructed enriched FCGs from Smali code, incorporating function types (system or third-party API) and permission requirements for each node, and used Graph Convolutional Networks (GCNs) to train malware classifiers on these graphs.
Similarly, Feng et al. \cite{feng20} focused on extracting features directly from the disassembled code sections in CGdroid.
This approach first uses hand-crafted features like the number of string constants and instructions for each node, and then utilizes a GNN to learn graph embeddings and an MLP for final classification.
Vinayaka and Jaidhar \cite{vinayaka21} further extended this concept by incorporating FCG's graph structural attributes (e.g., node degree) and non-graph features extracted from the disassembled functions, such as method attributes and opcode summaries.
Yumlembam et al.  \cite{yumlembam22} took a more general approach, modeling apps as local graphs where nodes denote APIs and co-occurring APIs in the same code block as edges.
The authors explored features like centrality measures, permissions, and intents from the manifest file.
Lo et al. \cite{lo22} leveraged PageRank~\cite{page1999}, in/out degree, and betweenness centrality values as node attributes.
DeepCatra utilized call traces, opcode features, and TF-IDF for critical API identification~\cite{wu23}.

Moving beyond basic features, some approaches explored learning node embeddings using GNNs and NLP techniques~\cite{catal21,gunduz22}.
Errica et al. \cite{errica21} leveraged Contextual Graph Markov Models to learn embeddings based on call graph structure and out-degree features, followed by classification with a neural network.
Cai et al. \cite{cai21} and Xu et al. \cite{xu21} have explored leveraging word embedding techniques to analyze Android apps.
Xu et al. \cite{xu21} used the Skip-gram algorithm to transform Android opcodes into vectors for analysis.
Beyond Android malware, graph-based representations combined with GNNs have also proven effective for vulnerability detection in compiled programs, where unified source and binary code property graphs enable robust analysis~\cite{irtiza25}. Similarly, enhanced control flow graph representations that capture exceptional interprocedural edges have been proposed for more precise static analysis of binaries~\cite{bockenek24}.

These approaches highlight the active research in utilizing FCGs for Android app security analysis.
Beyond dataset quality, ML-based malware detectors also face challenges from adversarial attacks, where malware authors can manipulate behavioral features to evade detection~\cite{digregorio24}.
{\bf However, a crucial limitation of most existing studies is the reliance on datasets that contain obsolete and/or duplicate APKs.}
This inflates the reported performance of classification approaches.
Our work addresses this limitation by introducing a new malware classification dataset, BCG, that consists of new and unique FCGs.
BCG will pave the way for {a true evaluation of the effectiveness} of the aforementioned FCG-based methods and open new directions in the field of ML-driven Android app security analysis.

\begin{table}[!t]

    \caption{Comparison of previous Android-based FCG datasets and BCG. 
    }
    \centering
    \setlength{\tabcolsep}{2pt}
    \renewcommand{\arraystretch}{1.1}
    \resizebox{\columnwidth}{!}{
        \begin{tabular}{ccccc}
            \hline
            \textbf{Dataset}                                                                         & \textbf{\# APKs}          & \textbf{\makecell{Collection                                                                                \\ Period}} & \textbf{\# Types} & \textbf{\makecell{Family \\ Info.}} \\ \hline
            Drebin                                                                                   & 5,560                     & 2010-2012                                     & N/A                   & \makecell{Yes (179,\\ no benign)} \\
            
            AndroZoo                                                                                 & 25M                       & Dynamic                                       & N/A                   & No                                                   \\
            CICAndMal2017                                                                            & 10,854                    & 2015-2017                                     & 5                     & No                                                   \\ 
            CICMalDroid                                                                              & 17,341                    & 2018                                          & 5                     & No                                                   \\ 
            MalNet                                                                                   & 1.2M                      & 2006-2021                                     & 47                    & Yes (696)                                            \\ 
            MalNet-Tiny                                                                              & 5,000                     & 2006-2021                                     & 5                     & Yes (5)                                              \\ 
\textbf{ BCG (our work)}             & {{\centering {10,057}  }} & {\centering    {2017-2025} }                  & {26}                  & Yes {(126)}                                          \\ \hline
        \end{tabular}
    }
    \label{tab:dataset_summary}
\end{table}

\subsection{Android-based FCG Datasets}
While numerous graph classification datasets exist for various fields like bioinformatics and social networks, options for cybersecurity, specifically malware detection using graph analysis, are scarce.
Most existing datasets in this domain are closed-source.
Fortunately, a few publicly available options like CICAndMal2017, CICMalDroid, AndroZoo, Drebin, and MalNet provide valuable resources for researchers analyzing Android malware through graph structures.

Given the dynamic nature of the Android ecosystem and the interest of malicious entities in releasing APKs, several FCG datasets have been collected and curated to study malware characteristics.
The Drebin dataset is one of the first in this direction, offering 5,560 malware apps (179 families) collected between 2010-2012 by MobileSandbox~\cite{arp14}, but it lacks benign samples.
Drebin provides summaries of each malicious APK using 10 features like permissions and intents.
While the Drebin dataset is valuable for multi-class classification (identifying the top-$k$ most frequent malware families), binary classification (malware vs. benign) requires collecting benign samples from other sources.
The AndroZoo dataset contains over 25 million APKs, mostly benign and with some malware verified through VirusShare~\cite{allix16}.
AndroZoo is constantly updated and provides 12 features per APK, which includes sha256, sha1, md5, apk\_size, dex\_size, dex\_date, pkg\_name, vercode, vt\_detection, vt\_scan\_date, markets, and added.
It primarily consists of benign APKs, and the fraction of malware APKs is 15.74\% (4 million APKs).
The Canadian Institute for Cybersecurity (CIC) released two Android app datasets for malware analysis: CICAndMal2017~\cite{lashkari18} and CICMalDroid~\cite{mahdavifar20}.
The former one has 10,854 APKs collected between 2015-2017 and categorized by malware type (benign, adware, ransomware, SMS, and riskware) and includes network traffic features.
The latter one boasts a larger collection, 17,341 APKs from 2018, with similar malware classifications.
It provides not only static features (permissions, intents) but also dynamic behavior data (system calls) and network traffic (PCAP format).
MalNet is a more recent dataset of 1.2 million FCGs extracted from AndroZoo APKs collected between 2006-2021~\cite{freitas20}.
The dataset is categorized into 47 types and 696 families.
It also offers a smaller version with 5,000 FCGs, MalNet-Tiny, for efficient experimentation.
    {Another related dataset, MalRadar, includes 4,534 APKs collected from 2014 to 2021 and provides information only at the family level~\cite{wang22}.}
More recently, MH-100K~\cite{mh100k} introduced a large collection of 101,975 Android samples with extracted features including 166 permissions, 24,417 API calls, and 250 intents.
    However, MH-100K does not provide function call graphs and only offers binary labels (malware/benign) without malware type or family classification, limiting its applicability for graph-based and fine-grained malware classification research.
    In the Windows malware domain, FCG-MFD~\cite{fcgmfd} recently introduced 100,000 samples (50,000 malware, 50,000 benign) with FCGs and metadata for 35 malware families collected between 2023-2024. While FCG-MFD demonstrates the value of FCG-based approaches for Windows PE malware, it targets a different platform than Android and thus cannot be directly compared with Android-specific datasets like BCG.

While these datasets are valuable, they suffer from two key issues:
\begin{enumerate}
    \item Most FCGs are obtained from {\bf obsolete APKs}, latest being only from 2021 (most of which are also benign). The complexity of both benign and malicious applications has drastically changed in recent years. For example, in the past it was simple enough to classify malware by determining if phone SIM card details were sent over the network. However, now benign applications do this for two-factor authentication and even for unique tracking of individuals for leaderboards in games.
    \item The current datasets often contain  many {\bf duplicate APKs}, packaged with trivial differences, which potentially inflates the classification performance (details in Section \ref{sec:unique_apks}).
\end{enumerate}

In this work, we address these limitations by constructing BCG, a new dataset with unique and recent APKs filtered with respect to various criteria, to enable a more robust testbed for FCG-based malware classification.
Those criteria include the minimum APK size to exclude very small applications, a minimum number of edges in the FCG to focus on complex functionalities, and multi-engine validation via VirusTotal \cite{virustotal} to guarantee high-confidence malware labels.
A detailed comparison between our work and existing datasets is presented in Table~\ref{tab:dataset_summary}.

\begin{figure*}[!t]
    \centering
    \includegraphics[width=.8\linewidth]{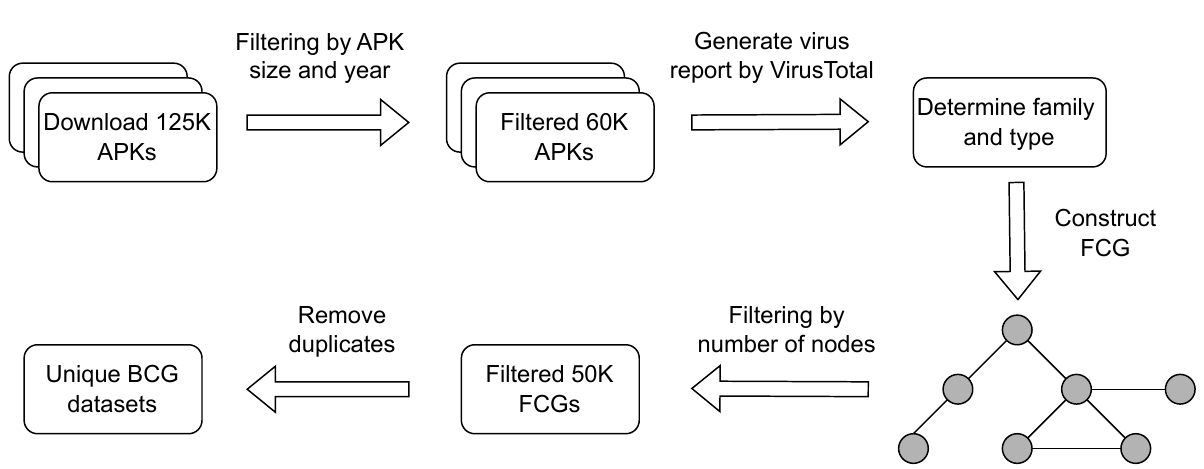}
    \caption{Construction process of BCG.}
    \label{fig:BCG_construction}
\end{figure*}

\section{Collecting and Filtering FCGs}
\label{sec:Dataset_Collection}

To ensure a robust and relevant BCG dataset, we constructed it by comprehensively analyzing both APK files and their corresponding graph properties.
This involved filtering and refining APKs based on various quality and relevance criteria, overcoming limitations of existing datasets.
There are some high-level observations
that guided our approach: (1) old APKs are simplistic in their structure and capabilities, (2) repackaging is very common, (3) certain virus families are over represented in datasets (often related to point 2), and (4) small APKs (based on bytecode size rather than auxiliary files) are often uninteresting from a detection standpoint.
A detailed flowchart of the BCG construction process is provided in Figure~\ref{fig:BCG_construction}. Here we summarize each step in this process:

\textbf{1. Downloading APKs:} \label{downloading_apks}
The foundation of our BCG dataset is built upon acquiring data sources that exhibit specific properties, including large and recent APKs (details of these properties are provided in Section~\ref{sec:BCG_properties}).
To achieve this, we secured permission from two prominent repositories: AndroZoo~\cite{allix16} and VirusShare~\cite{virusshare}.
{We selected AndroZoo and VirusShare as representative sources due to their extensive collections and widespread use in the research community. These are the de facto sources for the mobile, programming languages, and Android security communities. AndroZoo, for example, includes apps from the Google Play Store as well as other app sources such as AppChina and Anzhi (China-based App Store), encompassing a total of 25,412,146 APKs---significantly more than the number of apps currently available on the Google Play Store. This extensive repository covers all apps from these sources without any exclusions.}
We downloaded over {125,000}  APKs from the AndroZoo and VirusShare repositories.
    {Due to limitations in AndroZoo, specifically a restriction of 40 concurrent downloads and a cap of 500,000 APKs in a  six-month period per API key, we were unable to download all available APKs. Therefore, we prioritized downloading the most recent APKs first, followed by older ones in reverse chronological order. This approach led to fewer APKs from the years 2017 to 2020 (see detailed distribution in Figure \ref{fig:year_apk_count}). The entire download process took approximately 100 hours, and the resulting dataset required 2.5 TB of storage space.}
BCG covers APKs with DEX dates up to June 2025.

\textbf{2. Filtering APKs by Year and Size:}
To ensure our dataset has recent and large files, we followed a two-step filtering process on the {125,000} downloaded APKs.
Firstly, we extracted the DEX (Dalvik Executable) year from each APK. DEX year refers to the approximate year the application was compiled, and the information is embedded within the APK file.
By removing APKs with a DEX year before 2017, we ensured our dataset primarily reflects more recent applications.
Secondly, to ensure the APKs contained sufficient information for analysis, we removed any APKs with a size less than {1MB}.
We analyzed the distribution of APK size alongside the number of nodes and edges in their FCGs.
This analysis revealed that APKs exceeding {1MB} typically contained enough nodes and edges for meaningful FCG-based analysis.
Consequently, we filtered out APKs with a size less than  {1MB}.
These filtering processes resulted in a dataset of around {60,000} APKs.

\textbf{3. Determining Family and Type:}
To create the types and families in BCG, we utilized a multi-step process using VirusTotal \cite{virustotal} {and the AVClass package, aligning with the approach described in the MalNet \cite{freitas20} paper.}
First, we downloaded detailed reports for each APK using the VirusTotal API.
    {We then extracted virus categories and families from these reports using the AVClass package, which leverages results from 70 antivirus engines. AVClass assigns virus categories along with their counts based on these engines detections. We assigned a single label corresponding to the highest count. In cases where two labels tied for the highest count, we considered both and denoted them with $++$, following the approach used in the MalNet \cite{freitas20} paper.
        To ensure reliability, we considered APKs as malicious only if they were identified by at least two antivirus engines in AVClass. This criterion minimizes the risk of false positives caused by individual engine errors.      
        This threshold balances precision and recall: a single detection may represent a false positive from an overly aggressive engine, while requiring too many detections would exclude samples that only specialized engines can identify. The threshold of two follows prior work~\cite{freitas20} and provides a conservative estimate that minimizes false positives while retaining diverse malware samples.
        For benign samples, we confirm that none of the antivirus engines identified them as malware.}

\textbf{4. Constructing FCGs:}
Androguard has been effective in previous datasets for constructing FCGs \cite{desnos11}.
It conducts static analysis of the DEX file within each APK, identifying method names and their interactions.
Methods are represented as nodes, while calls between methods are depicted as directed edges.
We leverage Androguard to generate FCGs for each APK in our dataset.
To facilitate further analysis, we hash the method names into unique identifiers, enabling the creation of an integer edge list.
Specifically, we assign each unique method name a sequential integer ID within each graph, creating a consistent mapping that preserves the graph structure while enabling efficient storage and processing.
Our datasets are published in two formats: one containing the original method names as nodes, and another using hashed IDs.
    {The dataset also includes mappings from hashed IDs to original method names to facilitate traceability.
        These details allow future researchers to experiment with their own approaches to featurizing node-level (function-level) textual information.}

\textbf{5. Filtering APKs by Number of Edges:}
While the MalNet dataset has a large average graph size, it contains many very small graphs (less than 100 nodes/edges).
To exclude trivially small graphs (less than 100 nodes, as observed in MalNet), we further filtered our datasets to only include APKs with a minimum of 100 nodes in their FCGs, which are large enough to contain complex graph structure.
This filtering process resulted in a collection of around {50K graphs from 60K graphs.}

\textbf{6. Removing Duplicates:}
Existing datasets, such as MalNet and CICMalDroid, often contain duplicate APKs with identical functionalities (reflected in their FCG properties like number of nodes, edges, etc.) but disguised by different names. This can be due to repackaging (malware with minor changes) or the same APK appearing in multiple app stores.
To address this, we identified and removed duplicate FCGs based on six key graph properties of the FCGs: number of nodes, number of edges, average degree, in-degree centrality, size of largest connected component, and size of largest weakly connected component.
For further confirmation, we verified that these duplicates also shared the same malware type and family label.
Our methodology results in no false positives but there may well remain false negatives, i.e., undetected duplicates or FCGs that differ by one or a few nodes/edges, which is an interesting area for future work.
    {The FCGs with exactly identical values across all six properties are considered as duplicates and removed, resulting in a non-redundant final BCG dataset.}
    
\noindent {\bf Remark.} We chose exact matching on six properties as a conservative approach that prioritizes precision over recall in duplicate detection. Using approximate matching or fewer properties would identify more duplicates but risk removing genuinely distinct APKs. The six properties are selected for their discriminative power and computational efficiency; alternative metrics such as graph edit distance could capture near-duplicates but would significantly increase computational cost.

\section{Properties of BCG} \label{sec:BCG_properties}
While existing datasets like MalNet  \cite{freitas20}  offer valuable properties for evaluating Android malware with FCGs, they often contain duplicate APKs with different names but identical FCG structures.
Most of the previous datasets primarily consist of samples collected before 2017, potentially limiting their generalizability to modern malware and hence misleading ongoing research on malware detection.
Additionally, existing malware datasets often include numerous smaller-sized APKs, which limits their utility in comprehensive malware analysis.
Moreover, these datasets often lack essential APK properties, such as detailed information on the services or libraries used by the app, which impedes a thorough understanding of the app's behavior and functionality, making it difficult to accurately classify malware.
To address these limitations, we ensured that BCG has four key features: (1) larger graphs to facilitate more robust graph classification, (2) recent data (including 2017 and after) to reflect evolving threats, (3) unique APKs to ensure a more accurate evaluation testbed, and (4) non-graph APK features (APK attributes) for a more holistic evaluation. The summary of our datasets, BCG, is given in Table~\ref{tab:Dataset_table} and here, we briefly describe each of the four features:

\begin{table*}[!t]

\caption{Descriptive statistics for each FCG type in BCG. The table reports the number of graphs, unique malware families, and detailed statistics on node and edge counts per type (minimum, mean, maximum, and standard deviation). Labels marked with \texttt{++} (e.g., adware\texttt{++}trojan) indicate composite types assigned when two or more malware categories received equal detection counts from antivirus engines, as determined using AVClass and VirusTotal reports.}
\setlength{\tabcolsep}{4pt}
    \resizebox{\textwidth}{!}{
\centering
\begin{tabular}{@{}lcccccccccc@{}}
\toprule
\textbf{\makecell[l]{Final Label}} & \textbf{\makecell{\#\\Graphs}} & \textbf{\makecell{\#\\Fam.}} & \multicolumn{4}{c}{\textbf{\#Nodes}} & \multicolumn{4}{c}{\textbf{\#Edges}} \\
\cmidrule(lr){4-7} \cmidrule(lr){8-11}
& & & Min & Mean & Max & Std & Min & Mean & Max & Std \\
\midrule
benign & 5986 & 1 & 108 & 30.4K & 59.7K & 12.4K & 120 & 64.6K & 190.2K & 28.8K \\
trojan & 1757 & 90 & 101 & 22.7K & 59.5K & 13.5K & 109 & 50.5K & 143.2K & 31.0K \\
adware & 1438 & 98 & 101 & 22.1K & 58.7K & 13.0K & 110 & 49.0K & 159.9K & 30.6K \\
adware++trojan & 318 & 55 & 107 & 22.6K & 59.4K & 13.6K & 110 & 50.3K & 166.6K & 31.5K \\
trojan++riskware & 157 & 26 & 112 & 20.1K & 50.7K & 11.8K & 116 & 43.8K & 117.8K & 26.7K \\
adware++addisplay & 81 & 8 & 8069 & 27.2K & 45.4K & 7.2K & 18466 & 59.0K & 98.6K & 19.1K \\
riskware & 55 & 14 & 316 & 22.2K & 54.2K & 13.0K & 496 & 48.7K & 124.7K & 30.0K \\
adware++riskware & 54 & 11 & 115 & 15.1K & 40.4K & 12.6K & 147 & 34.4K & 97.2K & 29.1K \\
addisplay & 25 & 3 & 8627 & 26.6K & 39.4K & 8.0K & 15664 & 58.7K & 99.2K & 23.3K \\
trojan++downloader & 25 & 4 & 286 & 37.0K & 54.8K & 14.0K & 552 & 82.7K & 130.5K & 31.5K \\
trojan++addisplay & 20 & 4 & 8535 & 24.4K & 44.8K & 8.8K & 13151 & 53.7K & 92.1K & 20.9K \\
trojan++dropper & 17 & 9 & 107 & 25.6K & 54.0K & 17.2K & 153 & 55.8K & 115.8K & 40.6K \\
trojan++risktool & 14 & 5 & 232 & 22.3K & 50.5K & 14.6K & 374 & 45.3K & 115.9K & 32.0K \\
risktool & 13 & 5 & 6168 & 24.6K & 44.9K & 11.5K & 9887 & 55.7K & 105.7K & 29.1K \\
trojan++banker & 11 & 3 & 1723 & 19.8K & 50.7K & 13.9K & 3175 & 43.9K & 138.0K & 37.9K \\
adware++risktool & 10 & 3 & 19035 & 29.9K & 50.5K & 9.3K & 42336 & 66.8K & 116.8K & 22.2K \\
trojan++spr & 10 & 5 & 7575 & 21.7K & 36.1K & 10.1K & 10473 & 47.9K & 89.0K & 27.6K \\
spr & 9 & 2 & 16716 & 33.0K & 54.9K & 10.0K & 37396 & 75.6K & 130.8K & 24.5K \\
trojan++fakeapp & 9 & 2 & 9238 & 38.4K & 50.6K & 13.7K & 23753 & 80.2K & 104.4K & 25.3K \\
adware++spr & 8 & 6 & 232 & 26.8K & 48.0K & 16.4K & 374 & 63.5K & 112.8K & 40.6K \\
trojan++backdoor & 8 & 1 & 380 & 18.9K & 39.5K & 13.2K & 871 & 42.4K & 92.1K & 31.4K \\
adware++downloader & 7 & 1 & 126 & 34.2K & 49.6K & 23.5K & 156 & 69.3K & 101.0K & 47.7K \\
downloader & 7 & 2 & 983 & 27.2K & 47.5K & 21.5K & 1318 & 56.4K & 95.7K & 43.3K \\
fakeapp & 7 & 3 & 17053 & 37.4K & 53.9K & 17.2K & 35343 & 86.3K & 137.7K & 47.1K \\
trojan++clicker & 6 & 3 & 248 & 22.6K & 37.5K & 14.4K & 324 & 52.6K & 85.5K & 33.9K \\
riskware++risktool & 5 & 4 & 24090 & 30.7K & 37.5K & 6.6K & 47262 & 71.7K & 92.0K & 20.7K \\
\bottomrule
\end{tabular}
\label{tab:Dataset_table}
}
\end{table*}

\subsection{Larger Graphs}

Existing datasets for analyzing and differentiating APKs through graphs are often too small.
While MalNet has the largest dataset size, it contains many graphs with very few nodes/edges.
Out of 100K APKs in MalNet, approximately 3,000 FCGs contain less than 100 nodes/edges.
We address this limitation by creating a new dataset specifically designed for graph-based analysis. For this reason, we consider only those with a minimum size of {1MB} during the APK collection phase.
Additionally, we exclude any graphs with fewer than 100 nodes.
This ensures our dataset consists of larger graphs, providing more meaningful insights for malware classification as well as type/family classifications within malware.

\subsection{Recent and Modern APKs}

\begin{figure}[!htb]
    \centering
    \includegraphics[width=.97\linewidth]{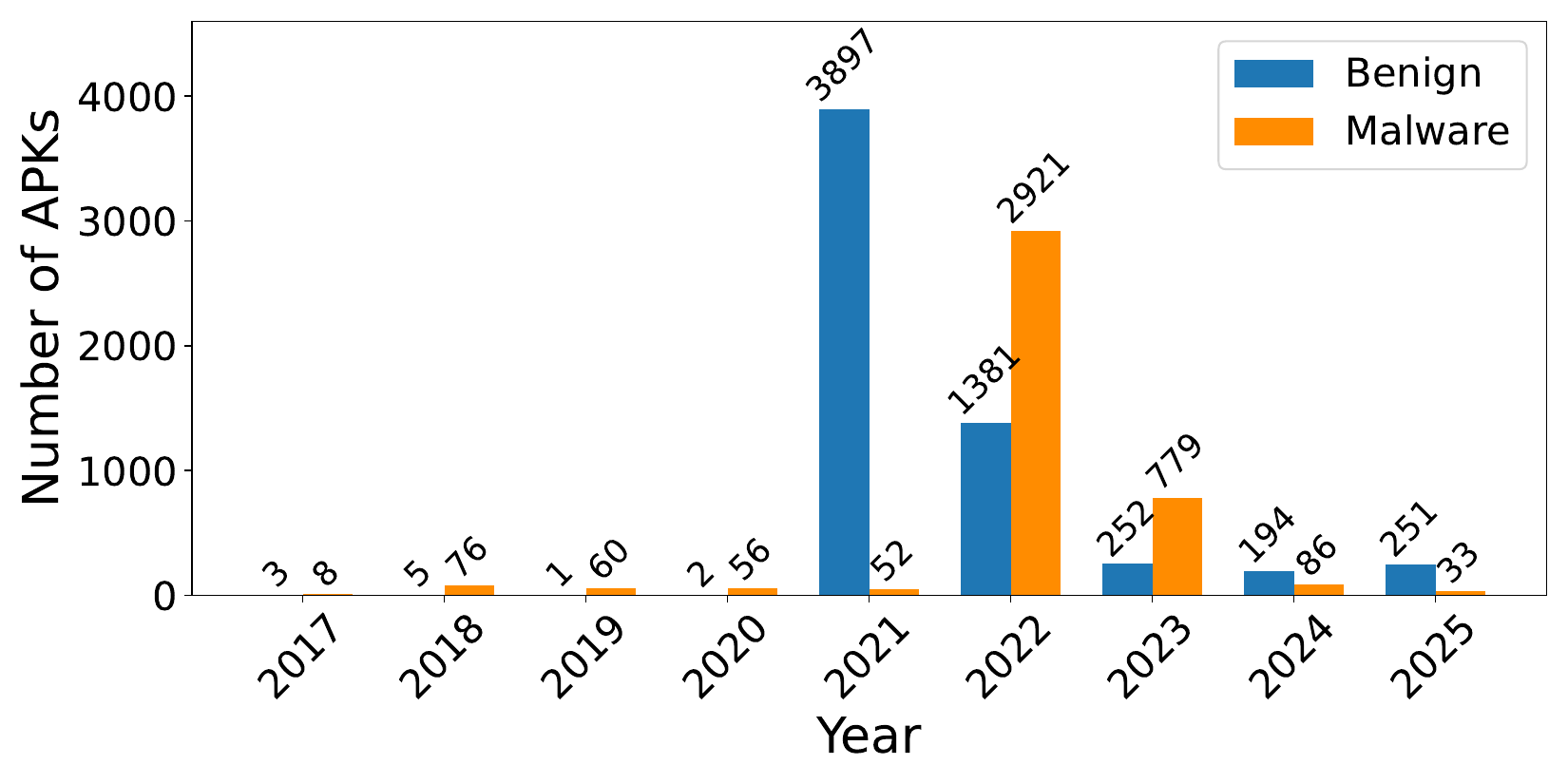}
     \vspace{-3ex}
    \caption{Temporal distribution of APKs in BCG.}
    \label{fig:year_apk_count}
    \vspace{2ex}
\end{figure}

All the existing FCG datasets contain mostly obsolete APKs.
For instance, approximately 99\% of the APKs in the MalNet dataset originate from before 2017, with only 1\% from post-2017.
Moreover, these post-2017 APKs are all benign.
The Android ecosystem has evolved significantly since 2016.
Older APKs, built for simpler Android versions, might not reflect the complexities of modern malware threats, which can limit the effectiveness of malware detection methods.
This phenomenon is closely related to concept drift in ML-based malware detection, where classifiers degrade as malware characteristics evolve over time~\cite{tripathi25}.
To address these limitations, we focus on constructing a new malware dataset that incorporates recent APKs.
We collected malware samples published in 2017 or later.
The distribution of APKs across different years is visualized in the histogram of Figure~\ref{fig:year_apk_count}.
{The exact counts and detailed distributions of type and family labels can be found in Table~\ref{tab:type_label_distribution} and Table~\ref{tab:family_label_distribution} in the Appendix.}
{We also include descriptions of unique malware types and the top 15 malware families in Table~\ref{tab:unique_malware_types_description} and Table~\ref{tab:top_15_malware_families_description}, respectively.}
It is evident that our dataset primarily contains recent APKs, with most APKs dating from 2021 onward.
    {Note that APK counts are lower in 2023–2025 because deduplication kept older APKs and discarded newer duplicates. Additionally, since we prioritized recent APKs and faced download limitations with AndroZoo, the number of APKs from 2017 to 2020 is relatively low (details in Section \ref{sec:Dataset_Collection}).}

\subsection{Unique APKs} \label{sec:unique_apks}

\begin{figure}[!t]
    \centering
    \subfloat[MalNet-Tiny]{\includegraphics[width=0.82\columnwidth]{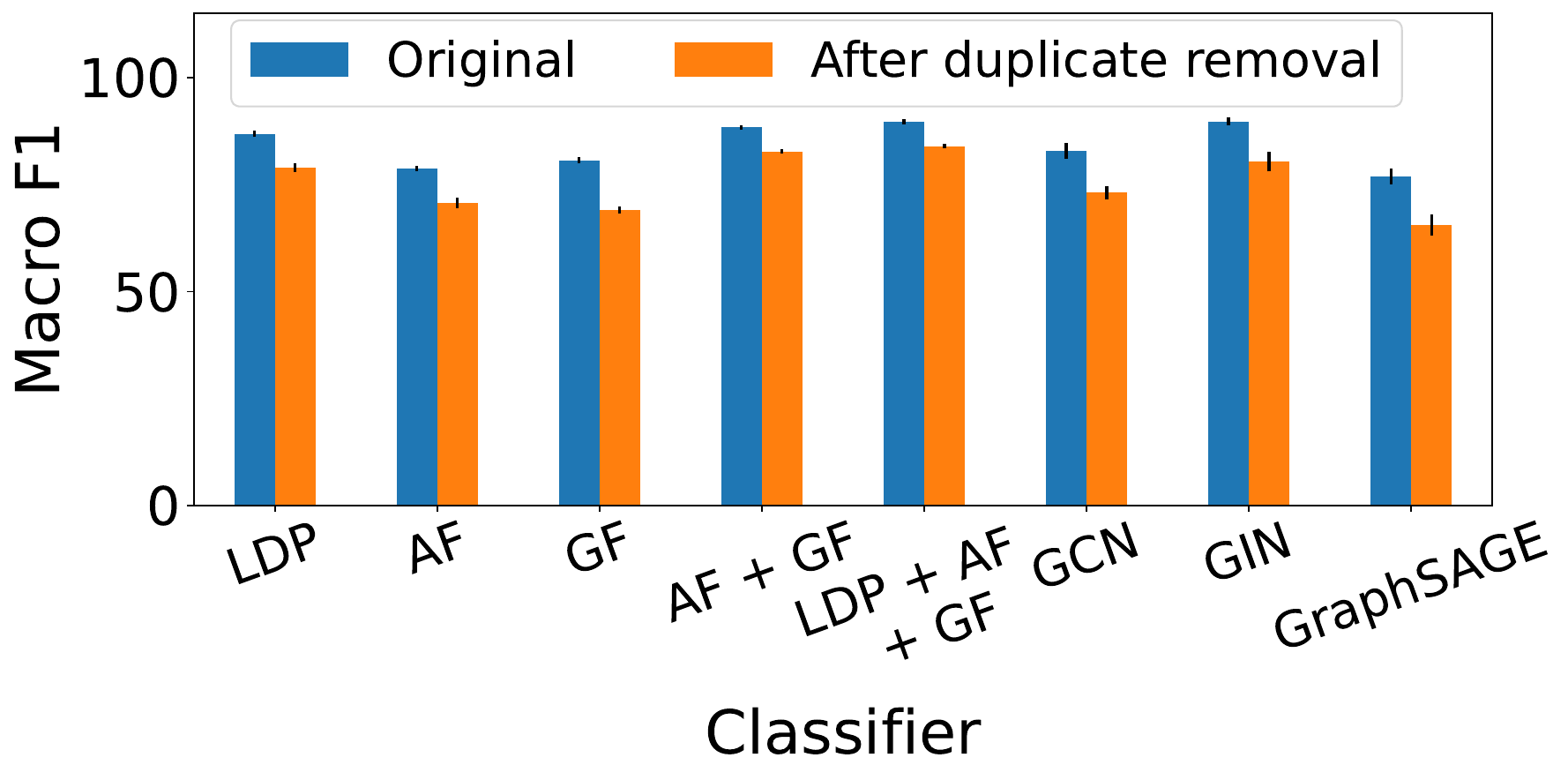}}\\[8pt]
    \subfloat[CICMalDroid]{\includegraphics[width=0.82\columnwidth]{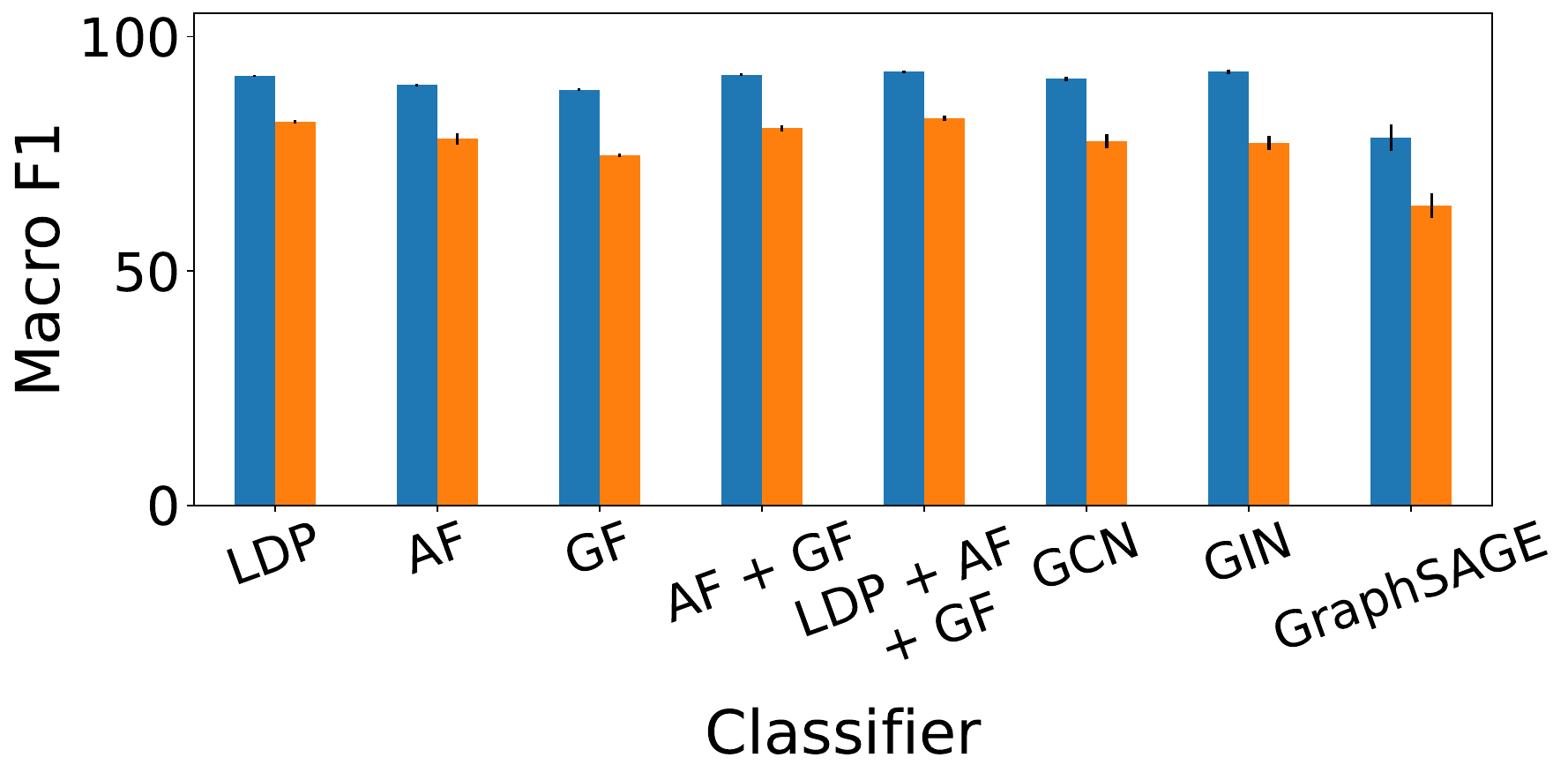}}
    \caption{Performance comparison across all methods on MalNet-Tiny and CICMalDroid datasets before and after removing the duplicate APKs. Removing duplicates decreases the performance drastically for all methods.}
    \label{fig:duplicate_results}
\end{figure}

High malware classification performance reported by recent approaches might be misleading due to the presence of duplicate APKs in commonly used datasets \cite{shirzad23,cao23,rampavsek22}.
The duplicate APKs have identical FCG structures but appear under different names.
This is caused by repackaging, where the same malware is redistributed with minor modifications like altered app icons or backgrounds, or by the same APK being uploaded to multiple app stores.
For instance, MalNet-Tiny and CICMalDroid suffer from this significantly.
MalNet-Tiny contains around 2,000 repackaged APKs (approximately 40\% of the entire data) while CICMalDroid has 41\% of duplicate APKs.
We have also investigated a subset of the MalNet, 100K out of 1.2M, and observed that approximately 51\% of the APKs are duplicates.
Importantly, removing these duplicates can often lead to a significant drop in the classification performance as the duplicate FCGs cause {\bf label leakage}, or even enable database lookups, in the original data when both train and test splits contain the same FCG.
We conducted experiments on the original MalNet-Tiny and CICMalDroid, as well as the filtered version after duplicates are removed, using different classifiers (details are in Section~\ref{sec:graph_classification_baselines}).
Figure~\ref{fig:duplicate_results} presents the results.
Macro F1 scores decrease after removing the duplicate APKs, consistently for all the classifiers, reaching up to 18.5\% relative decrease in CICMalDroid dataset with GraphSAGE.
Motivated by this, we focus on constructing new malware classification datasets that consist of unique APKs.

\begin{table}[!t]
    \caption{List of non-graph APK features and their descriptions.
    }
    \centering

        \begin{tabular}{lp{5.6cm}}
            \hline
            \textbf{Feature}   & \textbf{Description}                                                                                                                  \\
            \hline  
            APK size           & The size of the APK file in bytes.                                                                                                    \\ \hline
            DEX size           & The size of the Dex file in bytes.                                                                                                    \\ \hline
            App name           & The application name of the APK.                                                                                                      \\ \hline
            Package name       & The unique package name of the APK.                                                                                                   \\ \hline
            App permissions    & The list of permissions requested by the app, indicating the resources and data the app needs access to.                              \\ \hline
            App main activity  & The main activity of the app and  entry point when users launch the app.                                                              \\ \hline
            App all activities & The complete list of all activities defined in the app, representing the different screens and interactions available within the app. \\ \hline
            Services           & The list of all services used by the app, which are components that run in the background for long-running operations.                \\ \hline
            Receivers          & The list of all broadcast receivers in the app, which are components that respond to system-wide broadcast announcements.             \\ \hline
            Libraries          & The list of all libraries used by the app, including third-party ones that provide additional functionality and support.              \\

            \hline
        \end{tabular}
    \vspace{-1ex}
    \label{tab:APK_features}
\end{table}

\subsection{Non-graph APK Features (AF)} \label{sec:non-graph_apk_features}

Prior research \cite{wang20,lee19,alzaylaee20} highlights the effectiveness of basic APK features for malware classification.
However, existing datasets do not include these APK features, which require manual extraction from the APKs.
To address this, we have incorporated APK features into our BCG dataset.
We extracted two basic features from the APK and manifest files: APK size and DEX size.
APK size refers to the entire APK file size in bytes, while DEX size represents the size of the Dalvik Executable (DEX) file also in bytes.
The DEX file contains the optimized machine code used by the Android system to run the application.
Beyond basic size information, we utilize Androguard~\cite{desnos11} to extract various textual features from the APK, which have been shown to be useful for malware classification in previous works~\cite{wang20,lee19}.
These textual features include the app/package name, permissions requested by the app, all activities of the app, services or libraries used by the app, and the list of broadcast receivers.
Following prior work in malware classification~\cite{chen19,qiu20}, we encode each textual feature by training separate Word2Vec~\cite{mikolov13} models and computing 100-dimensional sentence embeddings via averaging word vectors. To enable efficient processing, we apply t-SNE~\cite{van08} to reduce the dimensionality of these embeddings to two.
    {t-SNE effectively preserves local structure and retains original information, and is widely adopted in prior malware detection studies \cite{yumlembam22,zhu18}. Therefore, we applied t-SNE to obtain two-dimensional features.}
t-SNE was applied after the train/test split; since this projection is lossy, AF serves as a simple baseline.
A detailed description of all non-graph APK features is provided in Table~\ref{tab:APK_features}.


\section{Experimental Evaluation} \label{sec:experimental_evaluation}
In this section, we evaluate the baseline performance of our dataset using various established methods.
Initially, we present the experimental setup and then we evaluate all the approaches.
The experiments were conducted on a Linux operating system (v. 3.10.0-1127) running on a machine with Intel(R) Xeon(R) Gold 6130 CPU processor at 2.10 GHz with 192 GB memory.
An Nvidia A100 GPU was used specifically for the GNN experiments.
    {\bf Our code is publicly available at {\url{https://github.com/erdemUB/BCG-code}}.}


\subsection{Experimental Setup}
To assess the graph classification performance of each model on a given dataset (i.e., set of graphs), we employ a 70/10/20 train/validation/test split.
We utilize macro-F1 score as the primary evaluation metric, considering the imbalanced nature of malware datasets, and also report accuracy, precision, and recall.
We repeat each experiment ten times with different random seeds and provide the average and standard deviation of these runs.

\subsection{Graph Classification Baselines} \label{sec:graph_classification_baselines}

We consider several state-of-the-art methods for classifying FCGs.
These methods encompass both feature-based approaches that analyze characteristics like permissions or app size directly from the APK, and graph-based approaches that focus on the FCG structure and incorporate node features for richer information.
Here we summarize each briefly:

\textbf{1. Deep Learning Techniques on Graphs:}
We consider three established methods based on GNNs and node embeddings: GCN \cite{kipf16B}, GIN \cite{xu18}, and LDP \cite{cai18}.
These methods, originally used for malware type and family classification on MalNet, are adapted to our setting.
LDP is a simple node representation scheme that summarizes each node and its immediate neighbors using five degree statistics.
These features are then aggregated and combined into feature vectors for the GNNs.
To ensure consistency, we adopt the same experimental setup in MalNet~\cite{freitas20}, using 5 GNN layers, Adam optimizer, 64 hidden units, a learning rate of 0.0001, and LDP node features for GCN and GIN.
We also incorporated GraphSAGE~\cite{hamilton17} into our evaluation alongside GCN and GIN given its established effectiveness for malware classification~\cite{lo22,yumlembam22,vinayaka21}.
Like GCN and GIN, GraphSAGE was implemented following the same experimental setup and identical hyperparameters across datasets for a consistent comparison.

\textbf{2. Random Forest on App-level Features:} \label{sec:app_level_features}
To investigate the effectiveness of basic APK properties that are unrelated to FCGs, we extracted various features from the original APK files by using Androguard (details are discussed in Section~\ref{sec:non-graph_apk_features}).
We refer to these as APK Features, AF for short.
We also constructed graph features, GF, from the FCGs, capturing simple graph analytics, such as the number of nodes/edges, largest connected component size, and centrality metrics; detailed descriptions are given in Table~\ref{tab:description_graph_features} in the Appendix.
Finally, we combine AF and GF, and feed them into a Random Forest model for malware classification.

\textbf{3. Combined Approach:} While both FCGs and basic APK features are valuable, recent research suggests that their combined use can lead to even better performance.
APK features capture high-level information about the app (permissions, size), while FCGs provide detailed insights into the app's functionality through call relationships between functions.
Hence, we explore the effectiveness of combining all app-level features (AF + GF) with LDP node embeddings derived from FCGs.
LDP node embeddings are aggregated to create graph-level feature vectors, which are then merged with AF + GF to form a comprehensive feature set.
We then evaluate this combined feature set using a Random Forest classification model.
To optimize hyperparameters like the number of estimators and tree depth, we perform a grid search on the validation set, replicating MalNet's configuration for these models.

\begin{table*}[!t]
    \centering

    \caption{Malware type classification results using different approaches on three datasets: MalNet-Tiny (MT), CICMalDroid (CMD), and our dataset BCG. MT* and CMD* indicate the results after eliminating duplicates. Best scores per dataset/method are in bold.
    }

    \setlength{\tabcolsep}{1.5pt}
    \resizebox{\textwidth}{!}{
        \begin{tabular}{|c|rrrrr||rrrrr|}
            \hline
            \multirow{2}{*}{\textbf{Method}}        & \multicolumn{5}{c||}{\textbf{Accuracy}}       & \multicolumn{5}{c|}{\textbf{Macro-F1}}                                                                                                                                                                                                                                                                                                                                                             \\ \cline{2-11}
                                                    & \multicolumn{1}{c|}{MT}                       & \multicolumn{1}{c|}{MT*}                      & \multicolumn{1}{c|}{CMD}                     & \multicolumn{1}{c|}{CMD*}                    & \multicolumn{1}{c||}{BCG} & \multicolumn{1}{c|}{MT}                       & \multicolumn{1}{c|}{MT*}                      & \multicolumn{1}{c|}{CMD}                       & \multicolumn{1}{c|}{CMD*}                    & \multicolumn{1}{c|}{BCG} \\ \hline
            {LDP}                                   & \multicolumn{1}{r|}{86.6 $\pm$ 0.7}           & \multicolumn{1}{r|}{77.9 $\pm$ 1.0}           & \multicolumn{1}{r|}{92.2 $\pm$ 0.1}          & \multicolumn{1}{r|}{87.3 $\pm$ 0.3}          & 75.2  $\pm$ 0.4           & \multicolumn{1}{r|}{86.7 $\pm$ 0.7}           & \multicolumn{1}{r|}{78.9 $\pm$  1.0}          & \multicolumn{1}{r|}{91.6 $\pm$ 0.1}            & \multicolumn{1}{r|}{81.8 $\pm$ 0.3}          & 18.3  $\pm$ 0.7          \\ \hline
            {AF}                                    & \multicolumn{1}{r|}{78.8 $\pm$ 0.6}           & \multicolumn{1}{r|}{74.8 $\pm$ 1.0}           & \multicolumn{1}{r|}{90.8 $\pm$ 0.2}          & \multicolumn{1}{r|}{85.0 $\pm$ 0.8}          & 74.8  $\pm$ 0.3           & \multicolumn{1}{r|}{78.7 $\pm$ 0.6}           & \multicolumn{1}{r|}{70.6 $\pm$ 1.2}           & \multicolumn{1}{r|}{89.7 $\pm$  0.0}           & \multicolumn{1}{r|}{78.1 $\pm$ 1.2}          & 11.9  $\pm$ 0.6          \\ \hline
            {GF}                                    & \multicolumn{1}{r|}{80.5 $\pm$ 0.7}           & \multicolumn{1}{r|}{66.8 $\pm$ 0.8}           & \multicolumn{1}{r|}{89.9 $\pm$ 0.2}          & \multicolumn{1}{r|}{82.3 $\pm$ 0.1}          & 70.5  $\pm$ 0.3           & \multicolumn{1}{r|}{80.6 $\pm$ 0.7}           & \multicolumn{1}{r|}{69.0 $\pm$ 0.8}           & \multicolumn{1}{r|}{88.6 $\pm$ 0.2}            & \multicolumn{1}{r|}{74.6 $\pm$ 0.3}          & 14.9  $\pm$ 0.3          \\ \hline
            {AF+GF}                                 & \multicolumn{1}{r|}{88.3 $\pm$  0.0}          & \multicolumn{1}{r|}{82.9 $\pm$ 0.5}           & \multicolumn{1}{r|}{92.6 $\pm$ 0.2}          & \multicolumn{1}{r|}{86.9 $\pm$ 0.4}          & 76.9  $\pm$ 0.3           & \multicolumn{1}{r|}{88.3 $\pm$  0.0}          & \multicolumn{1}{r|}{82.7 $\pm$ 0.5}           & \multicolumn{1}{r|}{91.9 $\pm$ 0.2}            & \multicolumn{1}{r|}{80.4 $\pm$ 0.6}          & 12.2  $\pm$ 0.4          \\ \hline
            {LDP+AF+GF}                             & \multicolumn{1}{r|}{89.6 $\pm$ 0.5}           & \multicolumn{1}{r|}{\textbf{83.8 $\pm$ 0.5}}  & \multicolumn{1}{r|}{\textbf{93.0 $\pm$ 0.2}} & \multicolumn{1}{r|}{\textbf{88.1 $\pm$ 0.3}} & \textbf{77.1  $\pm$ 0.5}  & \multicolumn{1}{r|}{89.6 $\pm$ 0.6}           & \multicolumn{1}{r|}{\textbf{83.9 $\pm$  0.0}} & \multicolumn{1}{r|}{\textbf{92.5 $\pm$ 0.2}  } & \multicolumn{1}{r|}{\textbf{82.5 $\pm$ 0.6}} & \textbf{18.8  $\pm$ 1.2} \\ \hline
            {GCN}                                   & \multicolumn{1}{r|}{82.8 $\pm$  1.0}          & \multicolumn{1}{r|}{73.2 $\pm$ 2.2}           & \multicolumn{1}{r|}{91.3 $\pm$ 0.4}          & \multicolumn{1}{r|}{84.9 $\pm$ 0.9}          & 72.6 $\pm$ 1.0            & \multicolumn{1}{r|}{82.8 $\pm$ 1.7}           & \multicolumn{1}{r|}{73.0 $\pm$ 1.5}           & \multicolumn{1}{r|}{91.0 $\pm$  0.0}           & \multicolumn{1}{r|}{77.7 $\pm$ 1.4}          & 11.3 $\pm$ 2.0           \\ \hline
            {GIN}                                   & \multicolumn{1}{r|}{\textbf{89.8 $\pm$  0.0}} & \multicolumn{1}{r|}{80.8 $\pm$ 1.9}           & \multicolumn{1}{r|}{92.7 $\pm$ 0.4}          & \multicolumn{1}{r|}{85.1 $\pm$ 0.5}          & 72.0 $\pm$ 1.5            & \multicolumn{1}{r|}{\textbf{89.7 $\pm$ 0.8}}  & \multicolumn{1}{r|}{80.3 $\pm$ 2.2}           & \multicolumn{1}{r|}{{92.5 $\pm$ 0.4}}          & \multicolumn{1}{r|}{77.3 $\pm$ 1.4}          & 10.0 $\pm$ 1.7           \\ \hline
            {GraphSAGE}                             & \multicolumn{1}{r|}{76.5 $\pm$  2.0}          & \multicolumn{1}{r|}{65.1 $\pm$ 2.2}           & \multicolumn{1}{r|}{79.1 $\pm$ 2.6}          & \multicolumn{1}{r|}{75.3 $\pm$ 2.4}          & 59.7 $\pm$ 3.6            & \multicolumn{1}{r|}{76.8 $\pm$ 1.8}           & \multicolumn{1}{r|}{65.5 $\pm$ 2.4}           & \multicolumn{1}{r|}{78.4 $\pm$ 2.8}            & \multicolumn{1}{r|}{63.9 $\pm$ 2.6}          & 6.0 $\pm$ 1.2           \\ \hline  \hline
            \multicolumn{1}{|l|}{\multirow{2}{*}{}} & \multicolumn{5}{c||}{\textbf{Precision}}      & \multicolumn{5}{c|}{\textbf{Recall}}                                                                                                                                                                                                                                                                                                                                                               \\ \cline{2-11}
            \multicolumn{1}{|l|}{}                  & \multicolumn{1}{c|}{MT}                       & \multicolumn{1}{c|}{MT*}                      & \multicolumn{1}{c|}{CMD}                     & \multicolumn{1}{c|}{CMD*}                    & \multicolumn{1}{c||}{BCG} & \multicolumn{1}{c|}{MT}                       & \multicolumn{1}{c|}{MT*}                      & \multicolumn{1}{c|}{CMD}                       & \multicolumn{1}{c|}{CMD*}                    & \multicolumn{1}{c|}{BCG} \\ \hline
            {LDP}                                   & \multicolumn{1}{r|}{87.4 $\pm$ 0.6}           & \multicolumn{1}{r|}{80.9 $\pm$ 0.8}           & \multicolumn{1}{r|}{91.3 $\pm$ 0.1}          & \multicolumn{1}{r|}{82.8 $\pm$ 0.5}          & 21.6  $\pm$ 0.6           & \multicolumn{1}{r|}{86.6 $\pm$ 0.7}           & \multicolumn{1}{r|}{78.6 $\pm$ 0.9}           & \multicolumn{1}{r|}{92.1 $\pm$ 0.3}            & \multicolumn{1}{r|}{\textbf{82.1 $\pm$ 0.4}} & 18.5  $\pm$ 1.0

            \\ \hline
            {AF}                                    & \multicolumn{1}{r|}{79.8 $\pm$ 0.5}           & \multicolumn{1}{r|}{75.5 $\pm$  2.0}          & \multicolumn{1}{r|}{89.2 $\pm$ 0.2}          & \multicolumn{1}{r|}{80.2 $\pm$  1.0}         & 13.0  $\pm$ 1.4           & \multicolumn{1}{r|}{78.8 $\pm$ 0.6}           & \multicolumn{1}{r|}{69.4 $\pm$ 1.0}           & \multicolumn{1}{r|}{90.4 $\pm$ 0.3}            & \multicolumn{1}{r|}{78.0 $\pm$ 1.1}          & 11.6  $\pm$ 0.4          \\ \hline
            {GF}                                    & \multicolumn{1}{r|}{81.1 $\pm$ 0.7}           & \multicolumn{1}{r|}{69.7 $\pm$ 1.1}           & \multicolumn{1}{r|}{88.5 $\pm$ 0.2}          & \multicolumn{1}{r|}{75.1 $\pm$  0.0}         & 15.3  $\pm$ 0.7           & \multicolumn{1}{r|}{80.5 $\pm$ 0.7}           & \multicolumn{1}{r|}{69.5 $\pm$ 0.7}           & \multicolumn{1}{r|}{88.8 $\pm$ 0.3}            & \multicolumn{1}{r|}{74.3 $\pm$ 0.3}          & 16.9  $\pm$ 0.3          \\ \hline
            {AF+GF}                                 & \multicolumn{1}{r|}{88.7 $\pm$ 0.5}           & \multicolumn{1}{r|}{83.9 $\pm$ 0.4}           & \multicolumn{1}{r|}{91.5 $\pm$ 0.2}          & \multicolumn{1}{r|}{82.4 $\pm$ 0.7}          & 13.5  $\pm$ 1.2           & \multicolumn{1}{r|}{88.3 $\pm$  0.0}          & \multicolumn{1}{r|}{82.6 $\pm$ 0.6}           & \multicolumn{1}{r|}{92.5 $\pm$ 0.2}            & \multicolumn{1}{r|}{79.9 $\pm$ 0.6}          & 12.1  $\pm$ 0.3          \\ \hline
            {LDP+AF+GF}                             & \multicolumn{1}{r|}{\textbf{90.1 $\pm$ 0.5}}  & \multicolumn{1}{r|}{\textbf{85.1 $\pm$  0.0}} & \multicolumn{1}{r|}{92.1 $\pm$ 0.3}          & \multicolumn{1}{r|}{\textbf{84.1 $\pm$ 0.7}} & \textbf{22.2  $\pm$ 2.1}  & \multicolumn{1}{r|}{89.6 $\pm$ 0.5}           & \multicolumn{1}{r|}{\textbf{83.9 $\pm$ 0.4}}  & \multicolumn{1}{r|}{\textbf{93.0 $\pm$ 0.2}}   & \multicolumn{1}{r|}{82.1 $\pm$ 0.5}          & \textbf{18.9  $\pm$ 1.0} \\ \hline
            {GCN}                                   & \multicolumn{1}{r|}{83.3 $\pm$ 1.4}           & \multicolumn{1}{r|}{72.8 $\pm$ 1.5}           & \multicolumn{1}{r|}{91.1 $\pm$ 0.5}          & \multicolumn{1}{r|}{77.4 $\pm$ 1.6}          & 13.3 $\pm$ 2.7            & \multicolumn{1}{r|}{82.8 $\pm$  1.0}          & \multicolumn{1}{r|}{74.7 $\pm$ 1.4}           & \multicolumn{1}{r|}{90.9 $\pm$ 0.5}            & \multicolumn{1}{r|}{78.3 $\pm$ 1.3}          & 11.9 $\pm$ 2.6           \\ \hline
            {GIN}                                   & \multicolumn{1}{r|}{{90.0 $\pm$ 0.7}}         & \multicolumn{1}{r|}{80.6 $\pm$ 2.2}           & \multicolumn{1}{r|}{\textbf{92.4 $\pm$ 0.5}} & \multicolumn{1}{r|}{77.8 $\pm$  0.0}         & 10.8 $\pm$ 2.1            & \multicolumn{1}{r|}{\textbf{89.7 $\pm$  0.0}} & \multicolumn{1}{r|}{81.2 $\pm$ 1.8}           & \multicolumn{1}{r|}{92.6 $\pm$ 0.4}            & \multicolumn{1}{r|}{77.6 $\pm$ 1.4}          & 10.6 $\pm$ 1.9           \\ \hline
            {GraphSAGE}                             & \multicolumn{1}{r|}{78.0 $\pm$ 1.6}           & \multicolumn{1}{r|}{65.2 $\pm$ 2.3}           & \multicolumn{1}{r|}{78.5 $\pm$ 2.6}          & \multicolumn{1}{r|}{64.3 $\pm$ 2.6}          & 6.1 $\pm$ 1.0            & \multicolumn{1}{r|}{76.5 $\pm$  2.0}          & \multicolumn{1}{r|}{67.5 $\pm$ 2.2}           & \multicolumn{1}{r|}{79.6 $\pm$ 2.4}            & \multicolumn{1}{r|}{64.1 $\pm$ 2.5}          & 6.7 $\pm$ 1.5           \\ \hline
        \end{tabular}
    }

    \label{tab:type_classification_results_all_datasets}
\end{table*}

\subsection{Performance Analysis} \label{sec:performance_analysis}

{\bf Malware Type Classification.} We evaluate the classification performance of the aforementioned classifiers on three datasets: MalNet-Tiny, CICMalDroid, and our new BCG dataset. Malware type (in BCG) belongs to one of 26 classes, either benign (not malware) or a specific malware type, as shown in Table~\ref{tab:Dataset_table}.
As reported in Table~\ref{tab:type_classification_results_all_datasets},
all the results on MalNet-Tiny and CICMalDroid are drastically better than those on BCG.
For accuracy, the best classifier yields $89.76\%$ and $93.04\%$ on MalNet-Tiny and CICMalDroid, whereas it is only {$77.1\%$} on BCG.
There is even a more drastic difference in Macro-F1 (and Precision and Recall): the best classifier can easily reach around $90\%$ on MalNet-Tiny and CICMalDroid but only $18.8\%$ on BCG!
Table~\ref{tab:type_classification_results_all_datasets} also includes the results of MalNet-Tiny and CICMalDroid datasets after duplicate removal.
Even after removing duplicates, all the methods achieve higher performance on these datasets than BCG.
Notably, the lowest Macro-F1 score (65.5) on MalNet-Tiny with GraphSAGE is significantly higher than the BCG equivalent {6.0}.
This highlights the difficulty of classifying malware in BCG.
We also compare against a majority-class baseline that always predicts benign, the largest class in BCG.
It attains 59.54\% accuracy but only 2.87\% Macro-F1 (2.29\% precision, 3.85\% recall) on the BCG test set.
All methods in Table~\ref{tab:type_classification_results_all_datasets} outperform it in Macro-F1, by 3 to 16 points; however, GraphSAGE's accuracy (59.7\%) is at the baseline's level, indicating that it largely predicts the majority class.

    {From those results, it is evident that state-of-the-art approaches can easily obtain high performance for malware classification on previously curated datasets; however, the same cannot be said for our carefully constructed BCG dataset.}
The state-of-the-art methods perform poorly on BCG, particularly in Macro-F1.
This suggests that FCG-based malware classification on recent, deduplicated data is considerably harder than results on existing datasets indicate.
There is a clear need for new graph classification techniques that can handle the complexities within the BCG.

\begin{figure}[!t]
    \centering
    \subfloat[MalNet-Tiny]{\includegraphics[width=0.49\columnwidth]{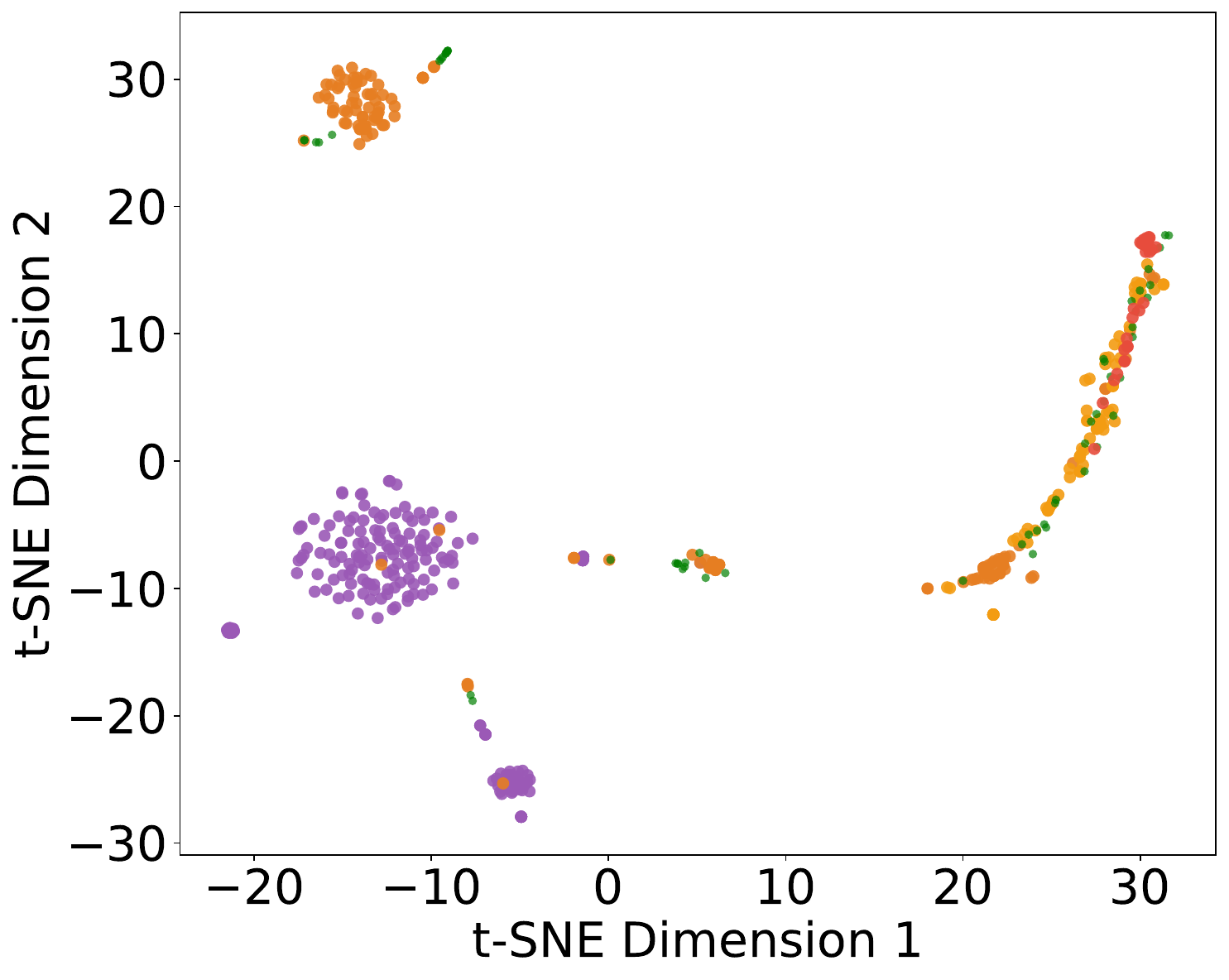}}\hspace{2pt}%
    \subfloat[BCG]{\includegraphics[width=0.49\columnwidth]{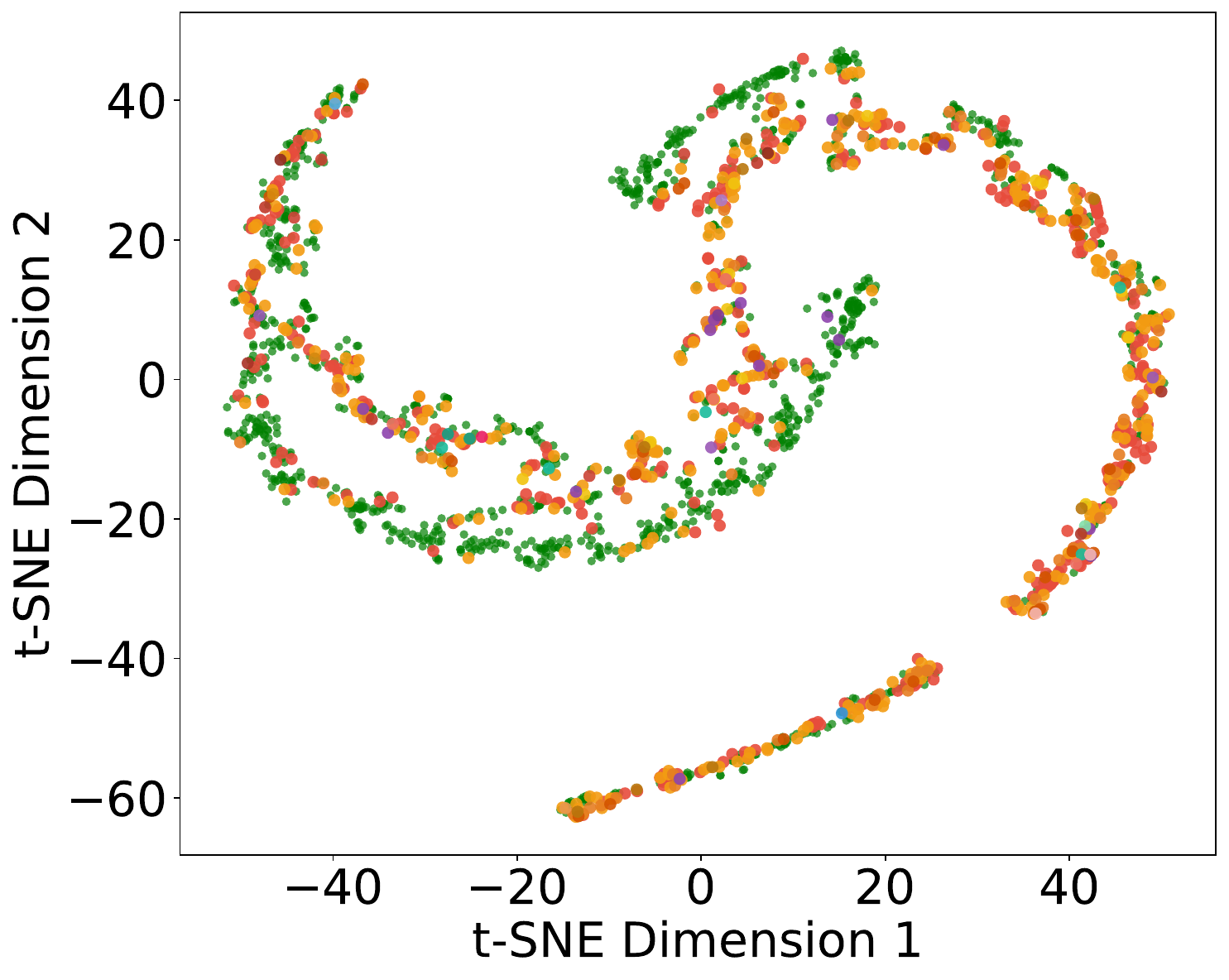}}
    \caption{t-SNE visualization of graph embeddings learned by GraphSAGE on MalNet-Tiny and BCG. Each color is a distinct malware or benign class.}
    \vspace{-10pt}
    \label{fig:tsne_graphsage}
\end{figure}

\begin{table}[!b]
    \vspace{-10pt}
    \centering
        \caption{Silhouette scores computed on model embeddings with true labels. Higher scores indicate better class separation. On MalNet-Tiny, all models achieve positive silhouette scores, reflecting clearer class boundaries in this dataset.}
    \setlength{\tabcolsep}{6pt}
        \resizebox{\columnwidth}{!}{
    \begin{tabular}{l|rrrr}
    \hline
    \textbf{Dataset} & \textbf{Raw LDP} & \textbf{GCN} & \textbf{GIN} & \textbf{GraphSAGE} \\
    \hline 
    BCG (26 Types)       & $-0.586$ & $-0.622$ & $-0.543$ & $-0.655$ \\
    MalNet-Tiny (5 Types) & $0.079$  & $0.191$  & $0.289$  & $0.225$  \\ \hline
\end{tabular}
}
    \vspace{5pt}
    \label{tab:silhouette_embed}
\end{table}

{{\bf Visualizing Model Embeddings.}}
To better understand the significant performance gap observed between BCG and MalNet-Tiny, we analyze and visualize the separability of the learned embedding spaces. Table~\ref{tab:silhouette_embed} presents the silhouette scores for four models and Figure~\ref{fig:tsne_graphsage} presents the t-SNE plots for GraphSAGE embeddings. MalNet-Tiny consistently achieves higher (positive) silhouette scores across models, indicating that samples belonging to the same class are grouped more compactly and are better separated from other classes in the learned embedding space. In contrast, BCG exhibits negative silhouette scores (though the differing number of classes, 26 vs.\ 5, limits direct comparability), suggesting substantial overlap among class representations, which makes it inherently more challenging for classification. This trend is also reflected in the t-SNE plots, where MalNet-Tiny forms more distinguishable and compact clusters, while BCG shows more dispersed and intermingled groupings. These observations provide further evidence that the structural similarity among graph types in BCG limits the ability of GNNs to learn well-separated representations, thereby contributing to the poor classification performance reported earlier.

\begin{table}[!t]
    \centering
        \caption{Malware family classification results on BCG dataset.}
    \setlength{\tabcolsep}{6pt}
    \resizebox{\columnwidth}{!}{
        \begin{tabular}{c|rrrr}
        \hline
            \textbf{Method} & \multicolumn{1}{c}{\textbf{Accuracy}}        & \multicolumn{1}{c}{\textbf{Macro-F1}}        & \multicolumn{1}{c}{\textbf{Precision}}       & \multicolumn{1}{c}{\textbf{Recall}} \\ \hline
            LDP             & \multicolumn{1}{r}{75.4 $\pm$ 0.37}          & \multicolumn{1}{r}{19.4 $\pm$ 0.85}          & \multicolumn{1}{r}{25.6 $\pm$ 1.01}          & 17.8 $\pm$ 0.83                      \\ 
            AF              & \multicolumn{1}{r}{76.6 $\pm$ 0.19}          & \multicolumn{1}{r}{16.0 $\pm$ 0.50}          & \multicolumn{1}{r}{21.8 $\pm$ 0.75}          & 14.7 $\pm$ 0.62                      \\ 
            GF              & \multicolumn{1}{r}{72.1 $\pm$ 0.55}          & \multicolumn{1}{r}{14.9 $\pm$ 1.26}          & \multicolumn{1}{r}{17.1 $\pm$ 1.71}          & 14.6 $\pm$ 1.18                      \\ 
            AF + GF         & \multicolumn{1}{r}{{79.6 $\pm$ 0.21}}        & \multicolumn{1}{r}{19.7 $\pm$ 0.37}          & \multicolumn{1}{r}{30.9 $\pm$ 0.65}          & 16.7 $\pm$ 0.33                      \\ 
            \parbox[c]{2.2cm}{\centering\vspace{2pt}LDP+AF+ GF\vspace{2pt}}
                            & \multicolumn{1}{r}{\textbf{79.9 $\pm$ 0.28}} & \multicolumn{1}{r}{\textbf{21.5 $\pm$ 1.01}} & \multicolumn{1}{r}{\textbf{33.1 $\pm$ 1.40}} & \textbf{18.0 $\pm$ 0.91}             \\ 
            GCN             & \multicolumn{1}{r}{63.6 $\pm$ 9.60}          & \multicolumn{1}{r}{1.16 $\pm$ 0.30}          & \multicolumn{1}{r}{1.05 $\pm$ 0.30}          & 1.23 $\pm$ 0.40                      \\ 
            GIN             & \multicolumn{1}{r}{71.7 $\pm$ 4.70}          & \multicolumn{1}{r}{3.46 $\pm$ 3.80}          & \multicolumn{1}{r}{3.61 $\pm$ 4.30}          & 3.02 $\pm$ 3.20                      \\ 
            GraphSAGE       & \multicolumn{1}{r}{68.5 $\pm$ 9.50}          & \multicolumn{1}{r}{0.85 $\pm$ 0.10}          & \multicolumn{1}{r}{0.63 $\pm$ 0.20}          & 0.84 $\pm$ 0.10                      \\ \hline
        \end{tabular}
    }
    \label{tab:BCG_family_results}
\end{table}

\begin{table}[!t]
\caption{Impact of temporal data split versus random partitioning on malware type classification (earlier data used for training and validation, later data for testing). }
\centering
\setlength{\tabcolsep}{3pt}
\resizebox{\columnwidth}{!}{
\begin{tabular}{c|rr|rr}
\hline
\multirow{2}{*}{\textbf{Method}}        & \multicolumn{2}{c|}{\textbf{Accuracy}}                               & \multicolumn{2}{c}{\textbf{Macro-F1}}                              \\ \cline{2-5} 
                                        & \multicolumn{1}{c|}{Random split}  & \multicolumn{1}{c|}{Time-based split} & \multicolumn{1}{c|}{Random split} & \multicolumn{1}{c}{Time-based split} \\ \hline
LDP                                     & \multicolumn{1}{r|}{75.24  $\pm$ 0.43}    & \textbf{59.39 $\pm$ 1.54}              & \multicolumn{1}{r|}{18.32  $\pm$ 0.65}   & \textbf{5.60 $\pm$ 0.13}                \\ 
AF                                      & \multicolumn{1}{r|}{74.84  $\pm$ 0.32}    & \textbf{51.82 $\pm$ 0.96}              & \multicolumn{1}{r|}{11.89  $\pm$ 0.62}  & \textbf{4.18 $\pm$ 0.27}               \\ 
GF                                      & \multicolumn{1}{r|}{70.47  $\pm$ 0.26}   & \textbf{54.55 $\pm$ 0.66}              & \multicolumn{1}{r|}{14.85  $\pm$ 0.34}   & \textbf{4.73 $\pm$ 0.16}               \\
AF + GF                                 & \multicolumn{1}{r|}{76.94  $\pm$ 0.27}    & \textbf{55.62 $\pm$ 0.73}              & \multicolumn{1}{r|}{12.23  $\pm$ 0.41}   & \textbf{4.93 $\pm$ 0.19}               \\
LDP + AF + GF                           & \multicolumn{1}{r|}{77.14  $\pm$ 0.53}    & \textbf{60.41 $\pm$ 1.18}              & \multicolumn{1}{r|}{18.80  $\pm$ 1.21}   & \textbf{5.86 $\pm$ 0.21}               \\ \hline \hline
\multicolumn{1}{l|}{\multirow{2}{*}{}} & \multicolumn{2}{c|}{\textbf{Precision}}                              & \multicolumn{2}{c}{\textbf{Recall}}                                \\ \cline{2-5} 
\multicolumn{1}{l|}{}                  & \multicolumn{1}{c|}{Random split}  & \multicolumn{1}{c|}{Time-based split} & \multicolumn{1}{c|}{Random split} & \multicolumn{1}{c}{Time-based split} \\ \hline
LDP                                     & \multicolumn{1}{r|}{21.63  $\pm$ 0.56}    & \textbf{6.21 $\pm$ 0.23}               & \multicolumn{1}{r|}{18.53  $\pm$ 1.01}   & \textbf{5.87 $\pm$ 0.13}               \\ 
AF                                      & \multicolumn{1}{r|}{12.99  $\pm$ 1.43}   & \textbf{4.62 $\pm$ 0.61}               & \multicolumn{1}{r|}{11.57  $\pm$ 0.41}  & \textbf{4.67 $\pm$ 0.19}               \\ 
GF                                      & \multicolumn{1}{r|}{15.26  $\pm$ 0.68}    & \textbf{4.85 $\pm$ 0.15}               & \multicolumn{1}{r|}{16.88  $\pm$ 0.30}   & \textbf{5.00 $\pm$ 0.14}                \\ 
AF + GF                                 & \multicolumn{1}{r|}{13.53  $\pm$ 1.17}    & \textbf{5.21 $\pm$ 0.22}               & \multicolumn{1}{r|}{12.06  $\pm$ 0.25}   & \textbf{5.11 $\pm$ 0.18}               \\ 
LDP + AF + GF                           & \multicolumn{1}{r|}{22.18  $\pm$ 2.08}    & \textbf{6.47 $\pm$ 0.30}               & \multicolumn{1}{r|}{18.92  $\pm$ 1.01}   & \textbf{6.04 $\pm$ 0.19}               \\ \hline
\end{tabular}
}
\label{tab:time_split_performance}
\end{table}

{\bf Malware Family Classification.}
Beyond malware type classification, our BCG dataset also includes {126} family labels over all APKs, enabling a more granular analysis.
We conducted experiments to classify malware according to their families, the results are in Table~\ref{tab:BCG_family_results}.
Interestingly, traditional GNNs like GraphSAGE (Macro-F1 $<$ 1\%) and GCN {(Macro-F1 = 1.16\%)} achieved low performance on this family classification task within the BCG dataset.
This suggests that these models might require further development or data augmentation to handle the complexities present in our data.

\begin{table}[!t]
\caption{Performance comparison of malware type classification between the first and second halves of BCG, with independent evaluation using separate train, validation, and test sets.}
\centering
\setlength{\tabcolsep}{3pt}
\resizebox{\columnwidth}{!}{
\begin{tabular}{c|rr|rr}
\hline
\multirow{2}{*}{\textbf{Method}}        & \multicolumn{2}{c|}{\textbf{Accuracy}}                               & \multicolumn{2}{c}{\textbf{Macro-F1}}                               \\ \cline{2-5} 
                                        & \multicolumn{1}{c|}{BCG first half}   & \multicolumn{1}{c|}{BCG second half} & \multicolumn{1}{c|}{BCG first half}   & \multicolumn{1}{c}{BCG second half} \\ \hline
LDP                                     & \multicolumn{1}{r|}{87.63 $\pm$ 0.21} & \textbf{63.39 $\pm$ 0.40}             & \multicolumn{1}{r|}{18.82 $\pm$ 0.31} & \textbf{16.34 $\pm$ 0.08}            \\ 
AF                                      & \multicolumn{1}{r|}{89.62 $\pm$ 0.23} & \textbf{62.69 $\pm$ 0.47}            & \multicolumn{1}{r|}{14.52 $\pm$ 0.66} & \textbf{11.23 $\pm$ 2.45}            \\ 
GF                                      & \multicolumn{1}{r|}{87.19 $\pm$ 0.34} & \textbf{55.82 $\pm$ 0.35}            & \multicolumn{1}{r|}{19.96 $\pm$ 1.04} & \textbf{13.09 $\pm$ 0.68}            \\ 
AF + GF                                 & \multicolumn{1}{r|}{89.81 $\pm$ 0.29} & \textbf{65.63 $\pm$ 0.35}            & \multicolumn{1}{r|}{14.53 $\pm$ 0.54} & \textbf{14.18 $\pm$ 1.77}            \\ 
LDP + AF + GF                           & \multicolumn{1}{r|}{89.75 $\pm$ 0.24} & \textbf{66.65 $\pm$ 0.62}            & \multicolumn{1}{r|}{19.77 $\pm$ 3.21} & \textbf{16.32 $\pm$ 0.34}            \\ \hline \hline
\multicolumn{1}{l|}{\multirow{2}{*}{}} & \multicolumn{2}{c|}{\textbf{Precision}}                              & \multicolumn{2}{c}{\textbf{Recall}}                                 \\ \cline{2-5} 
\multicolumn{1}{l|}{}                  & \multicolumn{1}{c|}{BCG first half}   & \multicolumn{1}{c|}{BCG second half} & \multicolumn{1}{c|}{BCG first half}   & \multicolumn{1}{c}{BCG second half} \\ \hline
LDP                                     & \multicolumn{1}{r|}{22.28 $\pm$ 0.41} & \textbf{19.99 $\pm$ 0.11}            & \multicolumn{1}{r|}{17.77 $\pm$ 0.25} & \textbf{15.41 $\pm$ 0.06}            \\ 
AF                                      & \multicolumn{1}{r|}{16.12 $\pm$ 1.47} & \textbf{11.27 $\pm$ 2.90}             & \multicolumn{1}{r|}{14.17 $\pm$ 0.49} & \textbf{11.58 $\pm$ 2.33}            \\ 
GF                                      & \multicolumn{1}{r|}{23.13 $\pm$ 1.24} & \textbf{14.18 $\pm$ 1.14}            & \multicolumn{1}{r|}{18.85 $\pm$ 0.88} & \textbf{13.88 $\pm$ 0.50}             \\ 
AF + GF                                 & \multicolumn{1}{r|}{15.60 $\pm$ 0.60}   & \textbf{14.16 $\pm$ 1.70}             & \multicolumn{1}{r|}{\textbf{14.15 $\pm$ 0.47} }& {14.45 $\pm$ 1.85}            \\ 
LDP + AF + GF                           & \multicolumn{1}{r|}{21.03 $\pm$ 3.25} & \textbf{21.01 $\pm$ 0.69}             & \multicolumn{1}{r|}{19.40 $\pm$ 3.23}  & \textbf{15.71 $\pm$ 0.35}            \\ \hline
\end{tabular}
}
\label{tab:half_performance}
\end{table}

{{\bf Evaluating the Difficulty of Classifying Recent APKs.}
The complexity of both benign and malicious applications has significantly evolved in recent years.
To empirically validate this, we conduct two sets of experiments.
For both experiments, we analyze results using Random Forest on app-level features and the combined feature set (as detailed in Section \ref{sec:graph_classification_baselines}). 
The first experiment involves a temporal split of the data, using earlier APKs for training and validation, and later APKs for testing.
Such a temporal split results in lower accuracy compared to random  partitioning as shown in Table \ref{tab:time_split_performance}.
This performance degradation is consistently observed across all feature configurations.
For example, the combined feature setting (LDP+AF+GF) shows a drop in accuracy from 77.14\% to 60.41\%, while Macro-F1 decreases from 18.80 to 5.86.
A similar trend is observed for AF, where accuracy drops from 74.84\% to 51.82\%, with Macro-F1 reducing from 11.89 to 4.18.
On average across all methods, accuracy drops from 74.93\% under random partitioning to 56.36\% with temporal splitting, while Macro-F1 decreases from 15.22 to 5.06.
Similar declines are observed in both precision and recall.
These results indicate that models trained on earlier APKs generalize poorly to more recent samples, suggesting increased classification difficulty over time.
For the second experiment, we divide the BCG dataset into two equal halves: {2017-April 2022 and May 2022-2025.}
Each half is independently evaluated using its own training and testing sets.
As shown in Table~\ref{tab:half_performance}, performance in the second half is consistently lower.
More specifically, when independently evaluated on temporally separated halves of the BCG dataset, all methods exhibit a marked reduction in performance in the more recent subset (May 2022–2025).
For instance, LDP+AF+GF accuracy decreases from 89.75\% in the first half (2017–April 2022) to 66.65\% in the second half.
Similarly, GF drops from 87.19\% to 55.82\% in accuracy, with Macro-F1 reducing from 19.96 to 13.09.
On average across all methods, accuracy decreases from 88.80\% in the first half to 62.84\% in the second half, with Macro-F1 decreasing from 17.52 to 14.23.
These consistent declines suggest that recent APKs are systematically harder to classify even under contemporaneous evaluation.
Taken together, both experimental settings show that classification performance decreases as the temporal gap between training and testing data increases.
These findings collectively suggest that malware classification becomes increasingly challenging for more recent APKs, and including obsolete APKs in an FCG dataset can result in inflated classification performance.
Note that the class composition also varies across years (e.g., 3,897 benign vs.\ 52 malware APKs in 2021), so distribution shift, besides recency, may contribute to this degradation; for instance, benign APKs make up 84.9\% of the first-half test set but only 32.7\% of the second-half test set (Table~\ref{tab:temporal_class_distribution} in the Appendix).
These findings align with recent work on concept drift in malware detection~\cite{tripathi25}, which demonstrates that ML-based malware classifiers require periodic retraining as threats evolve. Our temporal experiments are consistent with this drift phenomenon in the Android FCG domain.
}


\section{Conclusion}\label{sec:conc}
Traditional malware classification datasets struggle with redundancy, limited size, and outdated data, {particularly within the Android ecosystem}. These limitations hinder the development of effective models for detecting modern {Android} malware threats. This work addresses these issues by introducing the BCG dataset, a collection of recent and unique {Android-specific} FCGs extracted from recent application packages. BCG consists of 10,057 FCGs, with an average graph size of 27K nodes and 58K edges, representing both benign and malicious Android apps across 26 malware types and 126 families.
The analysis of BCG dataset revealed promising avenues for future research.
By overcoming the limitations of existing datasets, BCG paves the way for significant advancements in malware classification research.

    {\bf Limitations.} While the BCG dataset is valuable for FCG-based Android malware classification, it does not encompass other aspects of malware analysis, such as identifying malicious code or unpacking obfuscated content.
    {Expanding BCG to include these attributes could enable more comprehensive analysis.
        Additionally, future researchers may consider incorporating code-based textual embeddings as node features, though this requires additional steps such as decompilation and normalization.}

\textbf{Reproducibility Statement:}
All of our experimental results are reproducible.
The code for replicating the baseline results is publicly available at {\bf \url{https://github.com/erdemUB/BCG-code}}, with further details provided in Section~\ref{sec:experimental_evaluation}.
The dataset and its descriptions can be accessed at {\bf \url{https://erdemub.github.io/BCG-dataset}}.

\textbf{Ethical Considerations:}
All APK samples in our dataset were obtained from publicly available repositories with appropriate usage permissions. No personally identifiable information is present in the dataset. To mitigate bias, we included malware from diverse categories and families. The dataset is intended for research purposes only, and we discourage its use in production systems without proper auditing and safeguards against potential misuse.


\section*{Acknowledgments}
This work was supported by NSF awards 2107089, 2236789, and SHF-2512151, and used resources from the Center for Computational Research at the University at Buffalo~\cite{ccr}.

\bibliographystyle{IEEEtran}
\bibliography{reference}

\onecolumn
\appendix
\setlength{\intextsep}{6pt}

\begin{table}[H]
\caption{{Distribution of BCG types across the years 2017 to 2025. Labels marked with \texttt{++} (e.g., adware\texttt{++}trojan) indicate composite types assigned when two or more malware categories received equal detection counts from AVClass antivirus engines.}}
\centering
\footnotesize
\setlength{\tabcolsep}{5pt}
\renewcommand{\arraystretch}{1.2}
{
\begin{tabular}{|l|r|r|r|r|r|r|r|r|r|}
\hline
\makecell{\textbf{Type/} \\ \textbf{Year}}  & \textbf{2017} & \textbf{2018} & \textbf{2019} & \textbf{2020} & \textbf{2021} & \textbf{2022} & \textbf{2023} & \textbf{2024} & \textbf{2025} \\ \hline \hline
benign               & 3  & 5  & 1  & 2  & 3897 & 1381 & 252 & 194 & 251 \\ \hline
trojan               & 1  & 19 & 16 & 19 & 30   & 1314 & 302 & 44  & 12  \\ \hline
adware               & 4  & 37 & 26 & 21 & 17   & 988  & 315 & 26  & 4   \\ \hline
adware++trojan       & 0  & 8  & 4  & 5  & 0    & 229  & 58  & 4   & 10  \\ \hline
trojan++riskware     & 0  & 1  & 0  & 2  & 2    & 130  & 19  & 2   & 1   \\ \hline
adware++addisplay    & 0  & 5  & 7  & 4  & 1    & 62   & 2   & 0   & 0   \\ \hline
riskware             & 2  & 2  & 0  & 1  & 0    & 38   & 12  & 0   & 0   \\ \hline
adware++riskware     & 0  & 1  & 4  & 0  & 0    & 34   & 13  & 1   & 1   \\ \hline
addisplay            & 0  & 0  & 2  & 2  & 0    & 21   & 0   & 0   & 0   \\ \hline
trojan++downloader   & 0  & 0  & 1  & 0  & 0    & 6    & 15  & 2   & 1   \\ \hline
trojan++addisplay    & 0  & 3  & 0  & 0  & 0    & 17   & 0   & 0   & 0   \\ \hline
trojan++dropper      & 0  & 0  & 0  & 0  & 1    & 7    & 6   & 1   & 2   \\ \hline
trojan++risktool     & 0  & 0  & 0  & 0  & 0    & 13   & 1   & 0   & 0   \\ \hline
risktool             & 0  & 0  & 0  & 1  & 0    & 8    & 4   & 0   & 0   \\ \hline
trojan++banker       & 0  & 0  & 0  & 0  & 0    & 7    & 2   & 2   & 0   \\ \hline
trojan++spr          & 0  & 0  & 0  & 0  & 0    & 9    & 1   & 0   & 0   \\ \hline
adware++risktool     & 1  & 0  & 0  & 0  & 0    & 7    & 0   & 2   & 0   \\ \hline
spr                  & 0  & 0  & 0  & 0  & 0    & 6    & 3   & 0   & 0   \\ \hline
trojan++fakeapp      & 0  & 0  & 0  & 0  & 0    & 1    & 6   & 1   & 1   \\ \hline
trojan++backdoor     & 0  & 0  & 0  & 0  & 0    & 7    & 0   & 0   & 1   \\ \hline
adware++spr          & 0  & 0  & 0  & 0  & 0    & 3    & 5   & 0   & 0   \\ \hline
fakeapp              & 0  & 0  & 0  & 0  & 1    & 2    & 4   & 0   & 0   \\ \hline
downloader           & 0  & 0  & 0  & 0  & 0    & 1    & 5   & 1   & 0   \\ \hline
adware++downloader   & 0  & 0  & 0  & 0  & 0    & 2    & 5   & 0   & 0   \\ \hline
trojan++clicker      & 0  & 0  & 0  & 0  & 0    & 5    & 1   & 0   & 0   \\ \hline
riskware++risktool   & 0  & 0  & 0  & 1  & 0    & 4    & 0   & 0   & 0   \\ \hline
\end{tabular}
}
\label{tab:type_label_distribution}
\end{table}

\vspace{3ex}

\begin{table}[H]
\caption{Distribution of BCG family labels (top 15) across the years 2017 to 2025. {Family labels marked with \texttt{++} (e.g., presenoker\texttt{++}smsreg) indicate composite family types assigned when two or more malware families received equal detection counts from AVClass antivirus engines.} }

\centering
\footnotesize
\setlength{\tabcolsep}{5pt}
\renewcommand{\arraystretch}{1.2}
{
\begin{tabular}{|l|r|r|r|r|r|r|r|r|r|}
\hline
\makecell{\textbf{Family/} \\ \textbf{Year}}  & \textbf{2017} & \textbf{2018} & \textbf{2019} & \textbf{2020} & \textbf{2021} & \textbf{2022} & \textbf{2023} & \textbf{2024} & \textbf{2025} \\ \hline \hline
presenoker           & 0  & 9  & 6  & 6  & 5    & 406  & 11  & 0  & 0  \\ \hline
smsreg               & 0  & 2  & 0  & 2  & 3    & 353  & 63  & 2  & 1  \\ \hline
hqwar                & 0  & 1  & 0  & 0  & 0    & 172  & 4   & 1  & 0  \\ \hline
kuguo                & 1  & 8  & 16 & 6  & 1    & 104  & 9   & 0  & 0  \\ \hline
ewind                & 0  & 2  & 0  & 1  & 0    & 93   & 13  & 0  & 0  \\ \hline
igexin               & 0  & 9  & 7  & 7  & 0    & 56   & 20  & 0  & 0  \\ \hline
dnotua               & 0  & 2  & 1  & 0  & 3    & 70   & 9   & 0  & 0  \\ \hline
dowgin               & 0  & 0  & 1  & 0  & 3    & 46   & 9   & 0  & 0  \\ \hline
wapron               & 0  & 0  & 0  & 1  & 0    & 54   & 0   & 0  & 0  \\ \hline
mobidash             & 0  & 1  & 0  & 0  & 0    & 49   & 3   & 0  & 0  \\ \hline
hypay                & 0  & 0  & 0  & 0  & 0    & 44   & 1   & 0  & 0  \\ \hline
youmi                & 0  & 4  & 1  & 1  & 0    & 16   & 20  & 0  & 0  \\ \hline
hiddad               & 0  & 0  & 0  & 0  & 0    & 26   & 14  & 2  & 0  \\ \hline
presenoker++smsreg   & 0  & 0  & 1  & 0  & 0    & 36   & 1   & 0  & 0  \\ \hline
autoins              & 0  & 1  & 1  & 0  & 0    & 27   & 6   & 1  & 0  \\ \hline
\end{tabular}
}
\label{tab:family_label_distribution}
\end{table}

\vspace{3ex}

\begin{table}[H]
\caption{Class distribution in each temporal split of BCG: the time-based split (Table~\ref{tab:time_split_performance}) and the first/second-half splits (Table~\ref{tab:half_performance}). The 10 largest types are shown; the remaining 16 types are aggregated as Other.}
\centering
\footnotesize
\setlength{\tabcolsep}{5pt}
\renewcommand{\arraystretch}{1.2}
{
\begin{tabular}{|l|r|r|r|r|r|r|r|r|r|}
\hline
\multirow{2}{*}{\textbf{Type}} & \multicolumn{3}{c|}{\textbf{Time-based split}} & \multicolumn{3}{c|}{\textbf{First half}} & \multicolumn{3}{c|}{\textbf{Second half}} \\ \cline{2-10}
 & \textbf{Train} & \textbf{Val} & \textbf{Test} & \textbf{Train} & \textbf{Val} & \textbf{Test} & \textbf{Train} & \textbf{Val} & \textbf{Test} \\ \hline \hline
benign               & 4,569 & 307 & 1,110 & 3,010 & 424 & 854 & 1,196 & 173 & 329 \\ \hline
trojan               & 1,086 & 308 & 363 & 216 & 33 & 57 & 1,000 & 159 & 292 \\ \hline
adware               & 829 & 264 & 345 & 168 & 27 & 68 & 811 & 114 & 250 \\ \hline
adware++trojan       & 187 & 59 & 72 & 49 & 7 & 5 & 193 & 18 & 46 \\ \hline
trojan++riskware     & 108 & 27 & 22 & 21 & 3 & 7 & 86 & 13 & 27 \\ \hline
adware++addisplay    & 76 & 3 & 2 & 15 & 2 & 5 & 45 & 6 & 8 \\ \hline
riskware             & 32 & 11 & 12 & 8 & 2 & 2 & 32 & 4 & 7 \\ \hline
adware++riskware     & 33 & 6 & 15 & 10 & 0 & 0 & 32 & 3 & 9 \\ \hline
addisplay            & 24 & 1 & 0 & 3 & 1 & 2 & 11 & 2 & 6 \\ \hline
trojan++downloader   & 5 & 2 & 18 & 1 & 0 & 0 & 18 & 1 & 5 \\ \hline
Other (16 types)     & 90 & 17 & 54 & 18 & 4 & 6 & 96 & 10 & 27 \\ \hline
\hline
Total                & 7,039 & 1,005 & 2,013 & 3,519 & 503 & 1,006 & 3,520 & 503 & 1,006 \\ \hline
Benign (\%)          & 64.9 & 30.5 & 55.1 & 85.5 & 84.3 & 84.9 & 34.0 & 34.4 & 32.7 \\ \hline
\end{tabular}
}
\label{tab:temporal_class_distribution}
\end{table}

\begin{table}[H]
\centering
\caption{Descriptions of unique malware types.}
\footnotesize
\renewcommand{\arraystretch}{1.15}
\begin{tabular}{|l|p{13.2cm}|}
\hline
\textbf{Unique Type} & \textbf{Description} \\
\hline \hline
 {adware}     & Displays unsolicited advertisements to users, often bundled with legitimate apps to generate revenue. \\\hline
 {addisplay}  & Aggressively forces advertisement display, including pop-ups and overlays, disrupting normal app usage. \\\hline
 {backdoor}   & Allows attackers remote access to a device, enabling unauthorized control or data theft. \\\hline
 {banker}     & Focuses on stealing banking credentials by logging input during online transactions. \\\hline
 {clicker}    & Simulates user interactions like ad clicks to fraudulently generate ad revenue. \\\hline
 {downloader} & Downloads and installs additional malicious payloads or components from the internet. \\\hline
 {dropper}    & Installs or unpacks other malware embedded within itself upon execution. \\\hline
 {fakeapp}    & Disguises itself as a legitimate application to deceive users and perform malicious activities. \\\hline
 {risktool}   & Contains potentially dangerous features (e.g., rooting or monitoring) which may be misused. \\\hline
 {riskware}   & Legitimate software that poses security risks if misconfigured or exploited. \\\hline
 {spr}        & Short for "spreader"; facilitates the propagation of malware across systems or networks. \\\hline
 {trojan}     & Appears benign but performs malicious actions like spying or privilege escalation once installed. \\\hline
\end{tabular}
\label{tab:unique_malware_types_description}
\end{table}

\begin{table}[H]
\centering
\caption{Descriptions of top 15 malware families in BCG dataset.}
\footnotesize
\renewcommand{\arraystretch}{1.15}
\begin{tabular}{|l|p{13.2cm}|}
\hline
\textbf{Family} & \textbf{Description} \\
\hline \hline
presenoker & Known for aggressive adware behavior and unauthorized promotion of apps without user consent. \\
\hline
smsreg & Abuses SMS functionality to send premium messages or register users for paid services without their knowledge. \\
\hline
hqwar & A trojan family that may perform hidden operations like unauthorized data collection or command execution. \\
\hline
kuguo & Adware often bundled with games and apps, designed to display intrusive advertisements. \\
\hline
ewind & Disguises itself as legitimate software to install additional payloads and track user behavior. \\
\hline
igexin & Embeds backdoor components into applications to collect sensitive data and connect to remote servers. \\
\hline
dnotua & Engages in silent installation of apps and background downloads, often without user interaction. \\
\hline
dowgin & Ad-displaying malware that modifies device configurations and downloads unwanted components. \\
\hline
wapron & Primarily associated with sending unauthorized SMS messages to premium-rate numbers. \\
\hline
mobidash & Focused on advertising fraud, often integrates into third-party apps to monetize fake interactions. \\
\hline
hypay & Targets payment systems and user financial data, commonly embedded in fake or cloned apps. \\
\hline
hiddad & Hides in repackaged apps and displays ads outside the host application to evade detection. \\
\hline
youmi & Collects user data such as location and app usage, previously banned from app stores due to violations. \\
\hline
presenoker++smsreg & Combines adware behavior (presenoker) with SMS abuse (smsreg) to maximize monetization and stealth. \\
\hline
autoins & Performs automatic installation of apps or updates without user approval, often for click fraud or ad revenue. \\
\hline
\end{tabular}
\label{tab:top_15_malware_families_description}
\end{table}


\begin{table}[H]
\centering
\caption{List of graph features and their descriptions.}
\footnotesize
\renewcommand{\arraystretch}{1.15}
\begin{tabular}{|l|p{10cm}|}
\hline
\textbf{Feature} & \textbf{Description} \\
\hline \hline
Num Nodes & Number of nodes in the FCG. \\ \hline
Num Edges & Number of edges in the FCG. \\ \hline
Node degree & Degree of the nodes in the FCG. \\ \hline
Selfloop & Number of self-loops in the FCG. \\ \hline
Indegree & Indegree of the nodes in the FCG. \\ \hline
Closeness & Closeness centrality of the nodes in the FCG. \\ \hline
Num Cycle & Number of cycles in the FCG. \\ \hline
Large Conn & Size of the largest connected component (LCC). \\ \hline
Large Conn Ratio & Ratio of the LCC size to the total number of nodes. \\ \hline
Large Weak Conn & Size of the largest weakly connected component (LWCC). \\ \hline
Large Weak Conn Ratio & Ratio of the LWCC size to the total number of nodes. \\ \hline
Second Large Weak Conn & Size of the second largest weakly connected component (2nd LWCC). \\ \hline
Second Large Weak Conn Ratio & Ratio of the 2nd LWCC size to the total number of nodes. \\ \hline
Power Alpha & Alpha parameter of the power law fitted to the node degrees. \\ \hline
Power Sigma & Sigma parameter of the power law fitted to the node degrees. \\ \hline
\end{tabular}
\label{tab:description_graph_features}
\end{table}

\end{document}